\documentclass[leqno,12pt]{article}

\usepackage{a4wide}
\usepackage[]{amsfonts} \usepackage[]{amsmath}
\usepackage[]{amssymb} \usepackage[]{latexsym}
\usepackage{graphicx,color} \usepackage{amsthm}
\usepackage{mathrsfs} \usepackage{url}
\usepackage[space]{grffile} \usepackage{xifthen}
\usepackage{algorithm}
\usepackage{algpseudocode}

\errorcontextlines\maxdimen

\allowdisplaybreaks

\makeatletter
    \newcommand*{\algrule}[1][\algorithmicindent]{\makebox[#1][l]{\hspace*{.5em}\thealgruleextra\vrule height \thealgruleheight depth \thealgruledepth}}%
\newcommand*{\thealgruleextra}{}
\newcommand*{\thealgruleheight}{.75\baselineskip}
\newcommand*{\thealgruledepth}{.25\baselineskip}

\newcount\ALG@printindent@tempcnta
\def\ALG@printindent{%
    \ifnum \theALG@nested>0
        \ifx\ALG@text\ALG@x@notext
        \else
            \unskip
            \addvspace{-1pt}
            \ALG@printindent@tempcnta=1
            \loop
                \algrule[\csname ALG@ind@\the\ALG@printindent@tempcnta\endcsname]%
                \advance \ALG@printindent@tempcnta 1
            \ifnum \ALG@printindent@tempcnta<\numexpr\theALG@nested+1\relax
            \repeat
        \fi
    \fi
    }%
\usepackage{etoolbox}
\patchcmd{\ALG@doentity}{\noindent\hskip\ALG@tlm}{\ALG@printindent}{}{\errmessage{failed to patch}}
\makeatother

\newbox\statebox
\newcommand{\myState}[1]{%
    \setbox\statebox=\vbox{#1}%
    \edef\thealgruleheight{\dimexpr \the\ht\statebox+1pt\relax}%
    \edef\thealgruledepth{\dimexpr \the\dp\statebox+1pt\relax}%
    \ifdim\thealgruleheight<.75\baselineskip
        \def\thealgruleheight{\dimexpr .75\baselineskip+1pt\relax}%
    \fi
    \ifdim\thealgruledepth<.25\baselineskip
        \def\thealgruledepth{\dimexpr .25\baselineskip+1pt\relax}%
    \fi
    \State #1%
    \def\thealgruleheight{\dimexpr .75\baselineskip+1pt\relax}%
    \def\thealgruledepth{\dimexpr .25\baselineskip+1pt\relax}%
}

\numberwithin{equation}{section}

\DeclareMathOperator*{\argmin}{arg\,min}

\newcommand\Tstrut{\rule{0pt}{2.6ex}}

\begin{document}

\theoremstyle{plain}
\newtheorem{theorem}{Theorem}[section] \newtheorem*{theorem*}{Theorem}
\newtheorem{proposition}[theorem]{Proposition} \newtheorem*{proposition*}{Proposition}
\newtheorem{lemma}[theorem]{Lemma} \newtheorem*{lemma*}{Lemma}
\newtheorem{corollary}[theorem]{Corollary} \newtheorem*{corollary*}{Corollary}

\theoremstyle{definition}
\newtheorem{definition}[theorem]{Definition} \newtheorem*{definition*}{Definition}
\newtheorem{example}[theorem]{Example} \newtheorem*{example*}{Example}
\newtheorem{remark}[theorem]{Remark} \newtheorem*{remark*}{Remark}
\newtheorem{hypotheses}[theorem]{Hypotheses} \newtheorem{assumption}[theorem]{Assumption}
\newtheorem{notation}[theorem]{Notation} \newtheorem*{question}{Question}

\newcommand{\ds}{\displaystyle} \newcommand{\nl}{\newline}
\newcommand{\eps}{\varepsilon}
\newcommand{\bE}{\mathbb{E}}
\newcommand{\cB}{\mathcal{B}}
\newcommand{\cF}{\mathcal{F}}
\newcommand{\cA}{\mathcal{A}}
\newcommand{\cM}{\mathcal{M}}
\newcommand{\cD}{\mathcal{D}}
\newcommand{\cN}{\mathcal{N}}
\newcommand{\cL}{\mathcal{L}}
\newcommand{\cLN}{\mathcal{LN}}
\newcommand{\bP}{\mathbb{P}}
\newcommand{\bQ}{\mathbb{Q}}
\newcommand{\bN}{\mathbb{N}}
\newcommand{\bR}{\mathbb{R}}
\newcommand{\VIX}{\mbox{VIX}}
\newcommand{\erf}{\mbox{erf}}
\newcommand{\LMMR}{\mbox{LMMR}}
\newcommand{\cLcir}{\mathcal{L}_{\!_{C\!I\!R}}}
\newcommand{\rhor}{\raisebox{1.5pt}{$\rho$}}
\newcommand{\varphir}{\raisebox{1.5pt}{$\varphi$}}
\newcommand{\taur}{\raisebox{1pt}{$\tau$}}
\newcommand{\spx}{S\&P 500 }

\newcommand{\der}[3][]{\ifthenelse{\isempty{#1}}{\frac{\partial #2}{\partial #3}}{\frac{\partial^{#1} #2}{\partial #3^{#1}}}}

\newcommand{\jpzchange}[2]{{\color{blue}#2}}

\title{The Calibration of Stochastic-Local Volatility Models - An Inverse Problem Perspective}

\author{Yuri F. Saporito\thanks{Escola de Matem\'atica Aplicada (EMAp), Funda\c{c}\~ao Getulio Vargas (FGV), Rio de Janeiro, Brazil,
{\em yuri.saporito@fgv.br}} , Xu Yang\thanks{Instituto de Matem\'atica Pura e Aplicada (IMPA), Rio de Janeiro, Brazil, {\em xuyang@impa.br}} \  and  Jorge P. Zubelli\thanks{Instituto de Matem\'atica Pura e Aplicada (IMPA), Rio de Janeiro, Brazil, {\em zubelli@impa.br}}}

\maketitle

\abstract{
We tackle the calibration of  the so-called Stochastic-Local Volatility (SLV) model. This is the class of financial models that combines the local and stochastic volatility features and has been subject of the attention by many researchers recently.  More precisely, given a local volatility surface and a choice of stochastic volatility parameters, we calibrate the corresponding leverage function. Our approach makes use of regularization techniques from the inverse-problem theory, respecting the integrity of the data and thus avoiding data interpolation.
The result is a stable and robust algorithm which is resilient to instabilities in the regions of low probability density of the spot price and of the instantaneous variance.
We substantiate our claims with numerical experiments using simulated as well as real data. 
}

\section{Introduction}
\label{sec:introduction}

The search for parsimonious models that would capture the market-observed smile behavior in the implied volatility surface (IVS) is still one of the main research topics in Mathematical Finance. Among the different models that have been introduced, perhaps the two most important attempts are the Stochastic Volatility (SV) models, \cite{heston1993} and \cite{gatheral06}, and the Local Volatility (LV) model of \cite{dupire1994}. While SV models capture crucial stylized facts of the volatility dynamics, they cannot perfectly calibrate the IVS, especially for short maturities. On the other hand, the LV model was constructed to fit any arbitrage-free IVS. However, it has poor dynamical properties, see \cite{alexander2004}. A very important issue when considering these models is their calibration to the market-observed IVS; we forward the reader to \cite{albani2015,kilin2011,mikhailov2004} and references therein for different calibration methods of SV and LV models, individually.

The Stochastic-Local Volatility (SLV) model is able to combine the best aspects of each one  of such model classes, see \cite{labordereGuyon2011,lee2014,slvKlebaner2015}. In the present article we shall present a stable and effective method to calibrate the SLV model that consists of adapting the method proposed in \cite{Egger-Engl2005} and \cite{engl1996regularization} to the SLV framework.

Although, separately, the calibration of SV and LV models has been extensively discussed in the literature, to the best of our knowledge, there are three approaches to calibrate an SLV model: \cite{labordere2009, labordereGuyon2011} and \cite{slvKlebaner2015}. The first two are Monte Carlo based methods, while the last one relies on the numerical solution of a partial differential equation (PDE). Since the method we propose here is also based on PDEs, we will use as \textit{benchmark} the method presented \cite{slvKlebaner2015}. Additionally, \cite{slv_numerical_pde_2017} uses the same calibration idea as in this benchmark method, but considers an adjoint method to solve the related Fokker-Planck equation for the transition probability density, see Section \ref{sec:numerical_fp}. 

In order to exemplify our method, we consider two numerical exercises. One uses synthetic data generated from a known SLV model and the other uses real option data from an FX market. These examples corroborate to the theoretical conclusions of the comparison of the benchmark and our proposed method. In fact, we verify that our method is more robust against noise and more resilient to instabilities.

The paper is organized as follows. In Section \ref{sec:model_description}, we briefly describe the SLV model. The benchmark and proposed calibration procedures are outlined in Section \ref{sec:calibration}. Finally, in Section \ref{sec:numerical_example}, we test our method with synthetic and real FX data. 

\section{Model Description}
\label{sec:model_description}

The Stochastic-Local Volatility (SLV) model assumes that, under a risk-neutral measure, the spot price satisfies
\begin{align}\label{eq:slv_model}
\left\{\begin{array}{l}
dS_t = (r - d)S_t dt + \sqrt{V_t}L(t,S_t)S_t dW_t^S,\\ \\
dV_t = \kappa(m - V_t)dt + \xi \sqrt{V_t} dW_t^V,\\ \\
dW_t^SdW_t^V = \rho dt.
\end{array}\right.
\end{align}

The rates $r$ and $d$ are the risk-free interest rate and the dividend rate, respectively. In this version of the SLV model, we assume that the stochastic part of the volatility is following the Heston model, \cite{heston1993}. The parameters $\kappa$, $m$, $\xi$ and $\rho$ have the same interpretation as in the pure SV model. Moreover, notice that this SLV model simplifies to the Heston model when $L \equiv 1$. With respect to our proposed calibration procedure, the choice of the SV model could have been easily modified. For example, we could have considered the SABR model of \cite{sabr2002} or the Inverse Gamma model of \cite{langrene2016}. Additionally, it is fairly easy to extend the method presented here to deal with time dependent interest and dividend rates, as we consider in our numerical examples. However, for cleaner exposition we will consider constant rates.

The function $L$ is called the \textit{leverage function} and it plays a very important role in the model above. It is the ingredient that allows the model to perfectly calibrate the IVS seen in the market. In order to achieve this goal, the function $L$ must satisfy (see \cite{gyongy1986})
\begin{align}
\sigma_{loc}^2(t,S) = \bE[V_t L^2(t,S_t) \ | \ S_t = S] = L^2(t,S) \bE[V_t  \ | \ S_t = S],\label{eq:local_vol_leverage}
\end{align}
where $\sigma_{loc}$ is the local volatility function calibrated to the market, see Section \ref{sec:local_vol}. We define then 
\begin{align}
\Sigma(t,S) =  \bE[V_t  \ | \ S_t = S].\label{eq:Sigma}
\end{align}
It is important to notice that Equation (\ref{eq:local_vol_leverage}) is an implicit equation for $L$, since it is needed for the computation of $\Sigma(t,S)$. 

Note that the parameters of the SV part of the model may be (almost) freely chosen. Given a reasonable choice of parameters, choosing $L$ to satisfy Equation (\ref{eq:local_vol_leverage}) allows the model to fit any arbitrage-free IVS. The adjectives \textit{almost} and \textit{reasonable} used here refer to the fact that the SDE (\ref{eq:slv_model}) using formula (\ref{eq:local_vol_leverage}) for $L$ might not have a solution for certain choices of parameters, see Remark \ref{rmk:not_exist}.

\section{Calibration}
\label{sec:calibration}

In this section we shall discuss two different PDE techniques that can be applied to calibrate an SLV model, namely the \textit{benchmark} and our proposed method. For both, we will assume that the local volatility surface and the SV parameters of the model have been already computed. 

Notice now that we can rewrite Equation (\ref{eq:Sigma}) as
\begin{align}\label{eq:Sigma_p}
\Sigma(t,S) = \bE[V_t  \ | \ S_t = S] =  \dfrac{\int_0^{+\infty} V p(t,S,V) dV}{\int_0^{+\infty} p(t,S,V) dV},
\end{align}
where $p(t,\cdot, \cdot)$ is the joint density probability $(S_t,V_t)$ and solves the Fokker-Planck PDE:
\begin{align}\label{eq:fokker_planck}
\frac{\partial p}{\partial t} &+ \frac{\partial}{\partial S}\left((r-d) S p\right) + \frac{\partial}{\partial V}(\kappa(m-V) p)-\frac{1}{2}\frac{\partial^2}{\partial S^2}(V L^2(t,S) S^2 p) \\
&-\frac{1}{2}\frac{\partial^2}{\partial V^2}(\xi^2 V p)-\frac{\partial^2}{\partial S\partial V}(\rho \xi V L(t,S) S p)=0,\nonumber
\end{align}
with initial condition $p(0,S,V)=\delta(S-S_0)\delta(V-V_0)$, i.e. the Dirac mass at $(S_0, V_0)$. 

\begin{remark}[Existence of Solution for SDE (\ref{eq:slv_model})]\label{rmk:not_exist}

Using Equation (\ref{eq:Sigma_p}), we may rewrite the SDE (\ref{eq:slv_model}) as
\begin{align}\label{eq:slv_model_p}
\left\{\begin{array}{l}
dS_t = (r - d)S_t dt + \sqrt{V_t}\sigma_L(t,S_t)\sqrt{\dfrac{\int_0^{+\infty} p(t,S_t,V) dV}{\int_0^{+\infty} V p(t,S_t,V) dV}}S_t dW_t^S,\\ \\
dV_t = \kappa(m - V_t)dt + \xi \sqrt{V_t} dW_t^V,\\ \\
dW_t^SdW_t^V = \rho dt.
\end{array}\right.
\end{align}
This is called a \textit{McKean SDE}, since the diffusion coefficient depends on the law of $(S,V)$. The existence of solutions of this SDE is a very challenging problem, and outside the scope of this paper. For a discussion of this topic, see \cite{labordereGuyon2011} and \cite{slv_existence}. For our work here, we will assume that the SDE has a unique strong solution.

\end{remark}

\begin{remark}[Mixing Fraction]

Additional parameters could be considered in order to calibrate some exotic derivatives (e.g. Barrier or Asian options). In particular, given some fixed vol-of-vol, $\xi$, and correlation, $\rho$, one could define
$$\xi_\lambda = \lambda \xi \mbox{ and } \rho_\lambda = \lambda \rho,$$
for $\lambda \in [0,1]$. This parameter is usually called \textit{mixing fraction}, as it mixes the stochastic and local aspects of the volatility.  Notice that $\lambda = 0$ implies a pure LV model. Moreover, the parameters $\xi$ and $\rho$ could be taken as the calibrated parameters of a pure SV model. The goal is then to choose $\lambda$ in order to calibrate a given exotic derivative price. For instance, if we choose a \textit{down-and-out barrier Call option} with barrier $B$ and strike $K > B$, we could numerically solve the following PDE
\begin{align}
\der{P}{t} + (r-d)S\der{P}{S} &+ \frac{1}{2}vL^2(t,S)S^2\der[2]{P}{S} + \kappa(m-v)\der{P}{v} + \frac{1}{2}\lambda^2\xi^2v \der[2]{P}{v} \\
&+ \lambda^2 \xi \rho v L(t,S) S \frac{\partial^2 P}{\partial S \partial V} - rP = 0, \nonumber
\end{align}
for $S \in [B,+\infty)$, with $P(t,B,V) = 0$ and final condition $P(T,S,V) = (S - K)^+$. It is straightforward to consider $\lambda$ time-dependent.

\end{remark}

\begin{remark}[Monte Carlo Methods]

For multi-factor SV models, both methods described in this paper require a high dimensional PDE solver to numerically deal with the Fokker-Planck equation and therefore suffers from the curse of dimensionality. This issue would be circumvented using a Monte Carlo method. For instance, in \cite{labordere2009}, using the Markovian projection technique, an algorithm is proposed to calibrate the leverage function $L$. Additionally, in \cite{labordereGuyon2011}, the authors applied the McKean's particle method, and developed an algorithm to hybrid models, where the short-term rate and the volatility are modeled as diffusions. 
\end{remark}

\subsection{Numerical Aspects}

There are some common numerical aspects for both benchmark and our calibration procedures, and we will state them here. Firstly, since the methods considered here are based on finite difference methods for PDEs, we will consider discrete meshes for time, spot price and spot volatility. A (uniform) mesh for a variable $y$ depends on a choice for a finite lower bound $y_{\min}$, a finite upper bound $y_{\max}$ and a step size $\Delta y$. It is assumed that $N_y = (y_{\max} - y_{\min})/\Delta y \in \bN$. The mesh for $y$ is then
$$y_i = y_{\min} + i\Delta y, \mbox{ for } i=0,\ldots, N_y.$$
In our case, we will assume that $t_{\min} = S_{\min} = V_{\min} = 0$. We will use the sub-index $n$ for $t$, $i$ for $S$ and $j$ for $V$. One could surely use non-uniform meshes, but we will present the results here with uniform meshes for clearer illustration.

\subsubsection{Numerical Methods for the Fokker-Planck PDE}\label{sec:numerical_fp}

The Fokker-Planck PDE, shown in Equation (\ref{eq:fokker_planck}), will have to be numerically solved given the parameters of the  SV model and a fixed leverage function $L$. That is, discretizing the Fokker-Planck PDE with any chosen method, we will compute an approximation for $p(t_n, S_i, V_j)$. A sensible choice for the discretization method is of Alternating Direction Implicit (ADI) type, see \cite{houtFoulon2010}. Namely, in our numerical example, we consider the Douglas scheme, which was proposed in \cite{douglas1956}. Moreover, the choice of boundary conditions for the numerical method is also very important. We have chosen the zero flux condition, see for instance \cite{boundary_flux}.

Note that, by using an ADI method to solve for the Fokker-Planck equation, $p(t_n,\cdot,\cdot)$ would depend on $L(t_n,\cdot)$ and $L(t_{n-1},\cdot)$. However, the benchmark method assumes that $p(t_n,\cdot,\cdot)$ only depends on $L(t_{n-1},\cdot)$. For more details, see Appendix \ref{sec:app}.

In a different direction, one could consider adjoint methods to numerically solve the Fokker-Planck PDE as in \cite{slv_numerical_pde_2017}.

\subsubsection{Approximation of the Initial Condition}\label{sec:initial_cond}

The initial condition for our Fokker-Planck PDE is not well-behaved; it is a Dirac mass at the point $(S_0,V_0)$. In order to avoid numerical issues arising from this lack of smoothness, we consider a smooth approximation of this Dirac mass. Specifically, we use
a bivariate normal distribution with small variances to represent the initial density:
\begin{align}\label{eq:initial_p}
p(0,S,V)=\frac{1}{2\pi \sigma_S \sigma_V} \exp\left\{-\frac{1}{2\sigma_S^2}(S-S_0)^2 -\frac{1}{2\sigma_V^2}(V-V_0)^2 \right\}.
\end{align}
In our numerical experiment, we have used $\sigma^2_S = \sigma^2_V = 10^{-3}$. See, for instance, \cite{slvKlebaner2015} for details.

\subsubsection{Numerical Computation of the Local Volatility}\label{sec:local_vol}

The calibration of local volatility surfaces is an important inverse problem in Mathematical Finance. In \cite{dupire1994}, the author has proposed the local volatility model, in which the European options prices satisfy the PDE of the form 
\begin{eqnarray}
\frac{\partial C}{\partial T} +  (r-d)K\frac{\partial C}{\partial K}-\frac 12 \sigma_{loc}^2(T,K)K^2 \frac{\partial^2 C}{\partial K^2} + 
dC=0, \quad T > 0, K > 0 , \label{x1} 
\end{eqnarray}
with initial and boundary conditions given by 
\begin{eqnarray}
C(0, K) &=& (S_0 - K)^+ , \label{x1bc}\\
\lim_{K \rightarrow \infty} C(T,K) &=& 0, \nonumber \\
\lim_{K \rightarrow 0} C(T,K) &=& S_0, \nonumber
\end{eqnarray}
where $C = C(T,K)$ is the value of the European call option with expiration date $T$ and strike price $K$. The inverse problem of the local volatility model is that, given the options prices $\{C(T,K)\}_{T,K}$, we want to find a plausible local volatility surface, $\{\sigma_{loc}(T,K)\}_{T,K}$, which can  explain these options prices. Two of the challenges of this inverse problem are the ill-posedness, \cite{DSZ2012}, and the scarceness of the data of options prices, \cite{albani2015}. To solve an ill-posed inverse problem, one popular method is to use the Tikhonov regularization \cite{ADZ2016,AZ2014,DSZ2012,DZ2015,ADZ2017}. We will briefly introduce this regularization method in Section \ref{sec:proposed_method}. To solve the problem of the scarceness of the data, one possibility is to interpolate/extrapolate the data of options prices to all the locations of the mesh, \cite{kahale2005smile}. In this paper, however, we apply the  method discussed in \cite{albani2015,AAZ2017}, where we use a $P$ matrix to map the grid locations of the estimated options prices to those of real data.

\subsection{Benchmark Method}
\label{sec:benchmark_method}

In this section, we will describe the method proposed in \cite{renMadanQian2007} and further developed in \cite{slvKlebaner2015}, which is our benchmark method. From Equation (\ref{eq:local_vol_leverage}), we have
$$L(t,S) = \frac{\sigma_{loc}(t,S)}{\sqrt{\Sigma(t,S)}} = \sigma_{loc}(t,S)\sqrt{\dfrac{\int_0^{+\infty} p(t,S,V) dV}{\int_0^{+\infty} V p(t,S,V) dV}}.$$

The \textit{benchmark} calibration procedure is based on the equation above. As we have previously mentioned, this is an implicit equation for $L$, since $p$ depends on it. More precisely, the leverage function is initialized at $S_i$ as
\begin{align}
L^B_{0,i} := \sigma_{loc}(0,S_i)\sqrt{\dfrac{\sum_{j=0}^{N_V} p_{0,i,j} \Delta V}{\sum_{j=0}^{N_V} V_j p_{0,i,j} \Delta V}},\label{eq:initial_L}
\end{align}
with $p_{0,i,j}$ given by Equation (\ref{eq:initial_p}). We are using the superscript $B$ to denote that this is the leverage function computed by the benchmark method. Assuming we have computed $L^B$ at time $t_n$, we use the numerical method discussed in Section \ref{sec:numerical_fp} to solve the Fokker-Planck Equation (\ref{eq:fokker_planck}) from $t_n$ to $t_{n+1}$ with $L(t,S_i) = L^B_{n,i}$, for $t \in [t_n, t_{n+1}]$. Hence, we find an approximation for $p(t_{n+1}, S_i, V_j)$, which we will denote by $p_{n+1,i,j}^B$. Finally, we set
\begin{align}\label{eq:update_L}
L^B_{n+1,i} := \sigma_{loc}(t_{n+1},S_i)\sqrt{\dfrac{\sum_{j=0}^{N_V} p_{n+1,i,j}^B \Delta V}{\sum_{j=0}^{N_V} V_j p_{n+1,i,j}^B \Delta V}},
\end{align}
and repeat the procedure above.

\begin{algorithm}
\caption{Benchmark Algorithm of \cite{renMadanQian2007}}\label{b_algorithm}
\begin{algorithmic}[1]
\myState{Set the initial condition of $p_{0,i,j}$ and $L_{0,i}^B$ using Equations (\ref{eq:initial_p}) and (\ref{eq:initial_L}), respectively.}
\For{$n=0,1,2,\ldots,N_t-1$}
\myState{Set $L(t,S_i) = L_{n,i}^B$, for $t \in [t_n, t_{n+1}]$.}
\myState{Solve the Fokker-Planck PDE (\ref{eq:fokker_planck}) in $t \in [t_n, t_{n+1}]$.}
\myState{Update $L^B_{n+1,i}$ with Equation (\ref{eq:update_L}).}
\EndFor
\myState{\Return $L_{n,i}^B$ for $n=0,\ldots, N_t$ and $i = 0, \ldots, N_S$.}
\end{algorithmic}
\end{algorithm}

\subsection{Proposed Method}
\label{sec:proposed_method}

The problem under consideration is a classical example of an ill-posed inverse problem. We shall now provide some background on inverse problems in general and on our specific problem.

Ill-posed problems have been treated extensively in the literature since they are relevant in several fields, see \cite{vogel2002} and
references therein. 
Amongst the main techniques to address these problems, it is safe to say that one of the most well-known is the so-called {\it Tikhonov regularization}. It consists basically in transforming the problem under consideration, say that of trying to solve $F(x)=y$, into a minimization of the form 
\begin{equation*}
 \argmin{ || F(x) - y ||^2 +\alpha |||x - x_0|||^2} \mbox{, }
\end{equation*}
where $||\cdot||$ and $|||\cdot|||$ are two norms, and $x_0$ incorporates the {\em a priori} information
that will allow the regularization of the problem. By changing the scale factor $\alpha$ of the norm  $|||\cdot|||$, one would put more or less emphasis on such {\em a priori} information. 
The optimal choice of $\alpha$ is the subject of intense
investigation. Among the more well-known methods one can cite the discrepancy principle and the $L$-curve method, see \cite{vogel2002}.
Further, developments led to the use 
of other metrics (or more generally functionals instead of norms), see \cite{KaltScherzer2008} and references therein. 

Let $L_{n,i}$ be the leverage function at time $t_n$ and spot price $S_i$, where $n=0,1,\ldots,N_t$ and $i=0,1,\ldots,N_S$, computed using our proposed method described below. Then the density function $p(t_{n+1},\cdot, \cdot)$ is computed by the numerical method discussed in Section \ref{sec:numerical_fp} to solve the Fokker-Planck Equation (\ref{eq:fokker_planck}) from $t_n$ to $t_{n+1}$ with $L(t,S_i) = L_{n,i}$, for $t \in [t_n, t_{n+1}]$. We denote this approximation by $p_{n+1, i, j}$.
Define $\mathcal{G}_1$ the operator that associates a given $\{L_{n,i}\}_{i=0}^{N_S}$ to the corresponding approximation of this density: 
\begin{equation}
\{p_{n+1,i,j}\}_{i,j=0}^{N_S,N_V} =: \mathcal{G}_1(\{L_{n,i}\}_{i=0}^{N_S}). \label{density_p}
\end{equation}
The initialization $\{L_{0,i}\}_{i=0}^{N_S}$ will be discussed in the sequel.

Fix now $\{y_i\}_{i=0}^{N_S}$ and let $\mathcal{G}_2$ be the operator mapping a choice of leverage function equals $\{y_i\}_{i=0}^{N_S}$ to the local volatility function at time $t_n$ following Equation (\ref{eq:local_vol_leverage}),
\begin{align}\label{eq:update_L_prop}
\mathcal{G}_2(\{y_i\}_{i=0}^{N_S}, \{p_{n,i,j}\}_{i,j=0}^{N_S,N_V}) := \left\{ y_i\sqrt{\dfrac{\sum_{j=0}^{N_V} V_j p_{n,i,j} \Delta V}{\sum_{j=0}^{N_V} p_{n,i,j} \Delta V}} \right\}_{i=0}^{N_S}.
\end{align} 
Notice
\begin{align*}
\mathcal{G}_2(\{y_i\}_{i=0}^{N_S}, \{p_{n,i,j}\}_{i,j=0}^{N_S,N_V})  &=\mathcal{G}_2(\{y_i\}_{i=0}^{N_S}, \mathcal{G}_1(\{L_{n-1,i}\}_{i=0}^{N_S})) =: \mathcal{G}(\{y_i\}_{i=0}^{N_S}, \{L_{n-1,i}\}_{i=0}^{N_S})
\end{align*} 
i.e. $\mathcal{G}$ is the operator that takes $\{L_{n-1,i}\}_{i=0}^{N_S}$ and $\{y_i\}_{i=0}^{N_S}$ to the local volatility at time $t_n$. Therefore, in order to obtain the surface of the leverage function, we have to solve the following Tikhonov-type optimization problem for $n=1,2,\ldots,N_t$.
\small
\begin{align}
\{L_{n,i}\}_{i=0}^{N_S} := &\argmin_{\{y_i\}_{i=0}^{N_S}}\|\sigma_{loc}(t_n,\cdot)-\mathcal{G}(\{y_i\}_{i=0}^{N_S}, \{L_{n-1,i}\}_{i=0}^{N_S})\|^2_{\Gamma^{-1}} \label{tik} \\
&+\alpha_1\|\{y_i\}_{i=0}^{N_S}-\{L_{n-1,i}\}_{i=0}^{N_S}\|^2_{D_0^{-1}}+\alpha_2\|R_S \{y_i\}_{i=0}^{N_S}\|^2_{D_S^{-1}}, \nonumber
\end{align}
\normalsize
where $\Gamma$, $D_0$ and $D_S$ are chosen symmetric positive definite covariance matrices.  We define the vector norm $\|x\|_{C}=\sqrt{x^TCx}$ and $R_S$ is the matrix representing the finite-difference 
approximation 
of the linear operator $\partial_S$. The initial value $\{L_{0,i}\}_{i=0}^{N_S}$ is chosen by solving the minimization (\ref{tik}) with $L_{-1, i} = c$, for all $i=0,\ldots,N_S$, for some chosen constant.
\begin{algorithm}%
\caption{Proposed Algorithm}\label{p_algorithm}
\begin{algorithmic}[1]
\myState{Set the initial condition of $\{p_{0,i,j}\}_{i,j=0}^{N_S,N_V}$ using Equation (\ref{eq:initial_p}) and set $\{L_{-1,i}\}_{i=0}^{N_S}$ to a chosen constant.}
\For{$n=0,1,2,\ldots,N_t$}
\myState{Solve the minimization problem (\ref{tik}) for $\{L_{n,i}\}_{i=0}^{N_S}$.}
\myState{If $n<N_t$ solve the finite difference problem in (\ref{density_p}) for $\{p_{n+1,i,j}\}_{i,j=0}^{N_S,N_V}$.} 
\EndFor
\myState{\Return $L_{n,i}$ for $n=0,\ldots, N_t$ and $i = 0, \ldots, N_S$.}
\end{algorithmic}
\end{algorithm}

\section{Numerical Example}
\label{sec:numerical_example}

We will now compare the methods described in Sections \ref{sec:benchmark_method} and \ref{sec:proposed_method} within synthetic and real data examples. The following information is common to both cases:

\begin{table}[h!]
\begin{center}
 \begin{tabular}{l | c c c c}
\hline 
Variable & Lower Bound & Upper Bound & Fine Mesh & Coarse Mesh  \\
\hline
time & 0 & 1 & 0.001 & 0.025 \Tstrut\\
log-moneyness & -3 & 3 & 0.025 & 0.05\\
volatility & 0 & 1 & 0.005 & 0.01\\
 \hline
 \end{tabular}
\caption{Mesh parameters} \label{tab:mesh_parameters}
 \end{center}
\end{table}

The coarse mesh is the one used in the finite difference methods in our numerical examples below.

\begin{figure}[h!]
\centering
\includegraphics[width=0.5\textwidth]{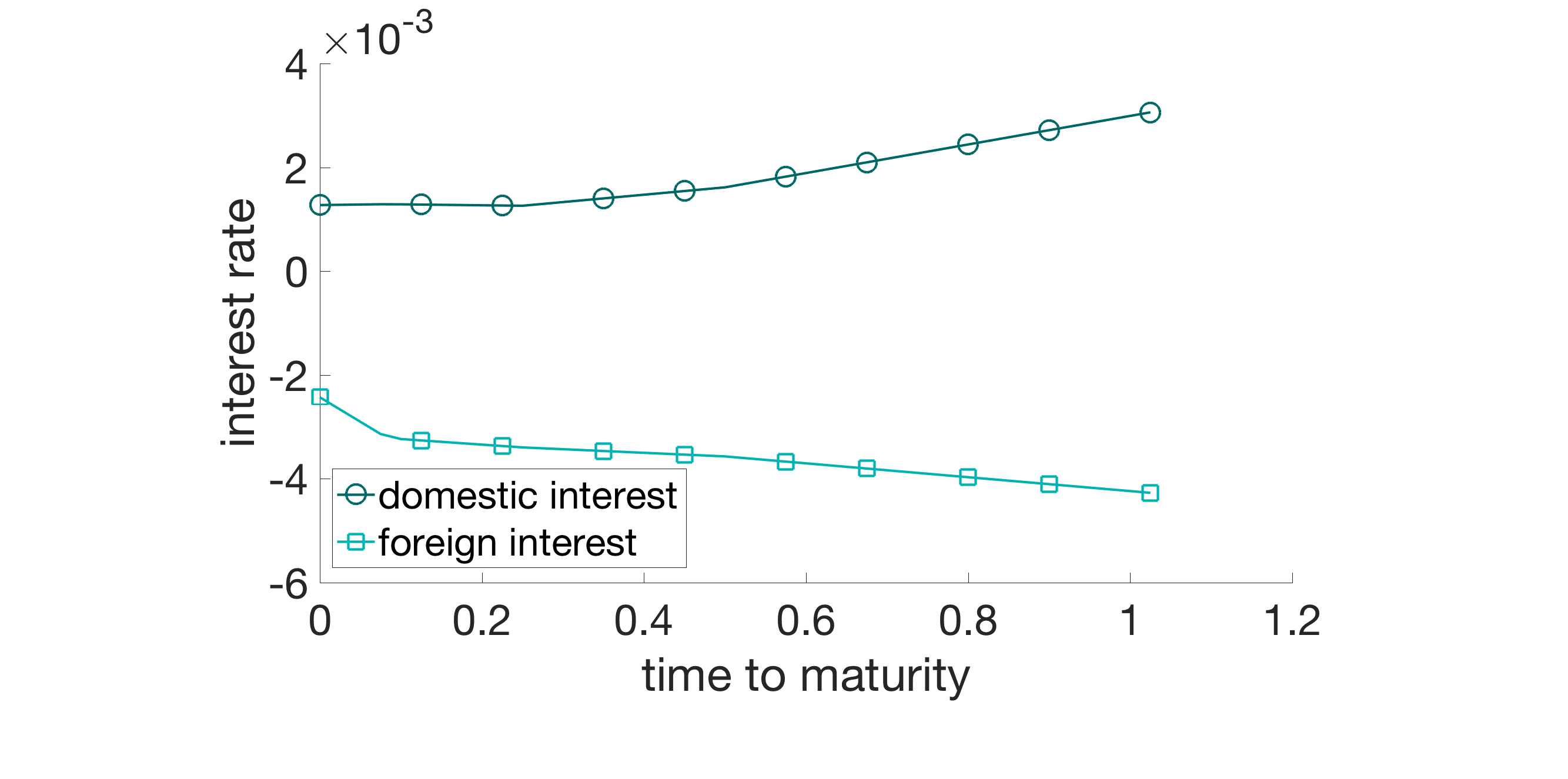}\hfill
\caption{Domestic and foreign interest rates}\label{fig:interest_rates}
\end{figure}

\subsection{Synthetic Data}\label{sec:syn}

\begin{figure}[H]
\centering
\includegraphics[width=0.5\textwidth]{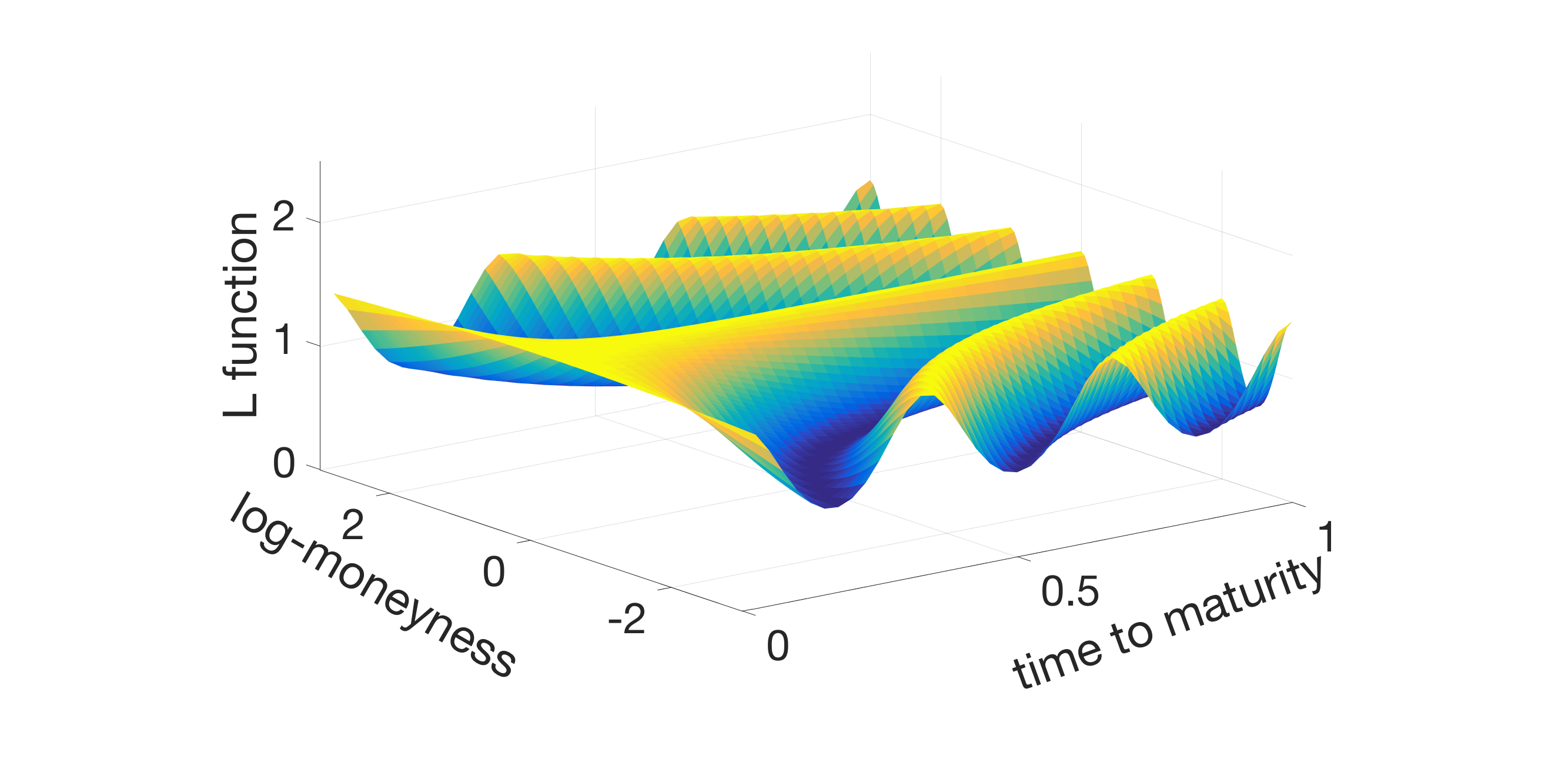}\hfill
\includegraphics[width=0.5\textwidth]{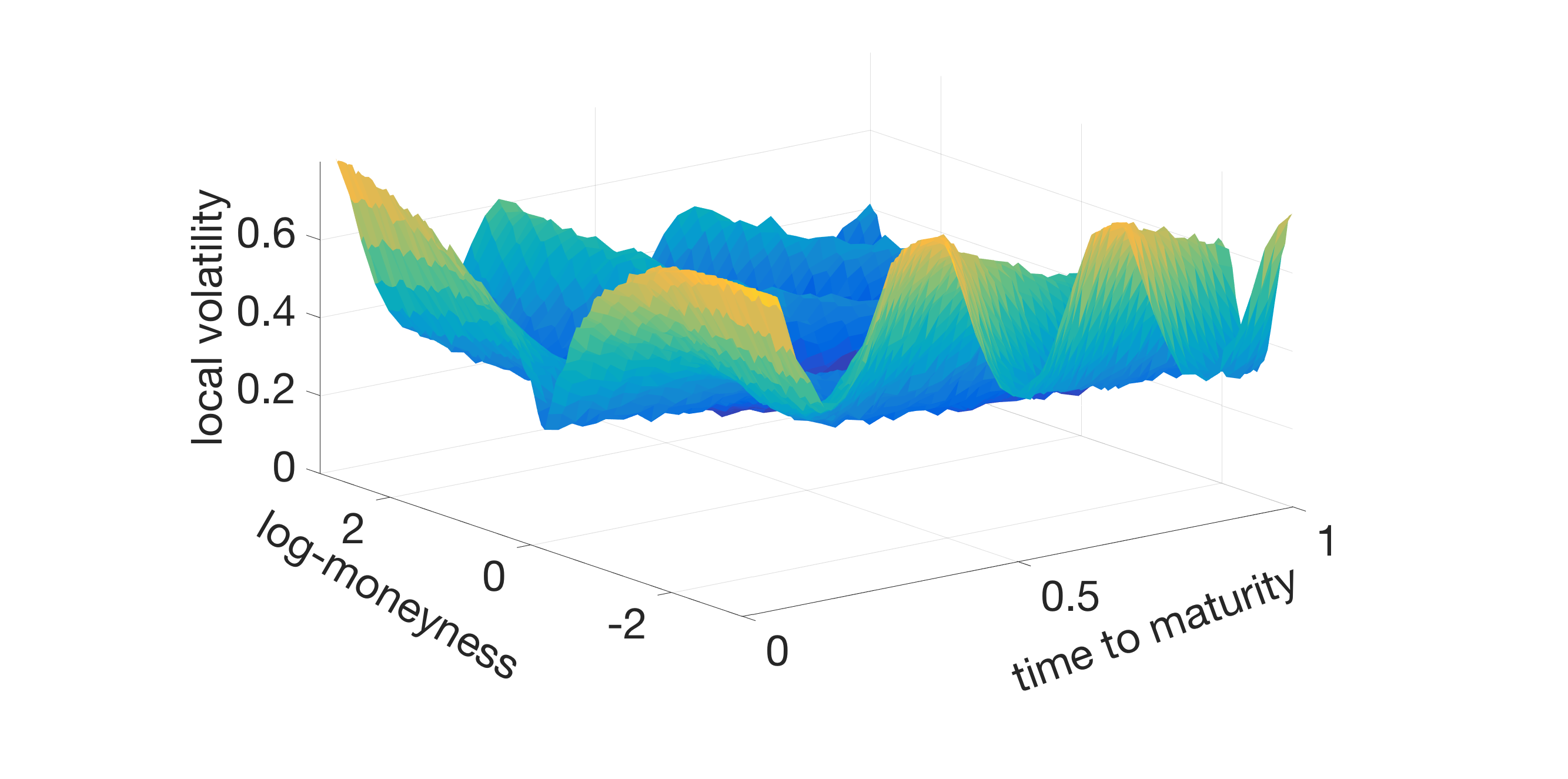}\hfill
\caption{The ground truth leverage function (left) and the  local volatility surface $\sigma_{loc}(t,x)^{\eta}$ (right).}\label{syn1}
\end{figure}

In this synthetic data example, we suppose the \textit{ground truth} leverage function (see Figure \ref{syn1}) is given by 
\begin{equation}
L(t,x):=1.1^{4\cos(2\pi xt)}, \mbox{ where } x\in [-3,3], t\in [0,1].\label{l_function}
\end{equation}
We calculate $\Sigma(t,x)$ and local volatility surface $\sigma_{loc}$ based on this given $L$ in the fine mesh. The details of the mesh for maturity, log-moneyness and volatility are given in the Table \ref{tab:mesh_parameters}. We then add a relative noise to the local volatility surface 
\begin{equation}
\sigma_{loc}(t,x)^{\eta}:=\sigma_{loc}(t,x)(1+0.01\eta_{t,x})
\end{equation}
where $\eta_{t,x}$ are independent draws from the standard normal distribution $\mathcal{N}(0,1)$. In order to avoid the so-called \textit{inverse crime} (\cite{kaipio2006statistical}), we sample the data to a coarser mesh, which is also given in Table \ref{tab:mesh_parameters}. Figure \ref{syn1} presents the noisy local volatility surface. The parameters of the SV part of the model are given in Table \ref{tab:syn_sv_parameters}. The Tikhonov parameters are $\alpha_1=0$ and $\alpha_2=10^{-2}$, see Equation (\ref{tik}).

\begin{table}
\begin{center}
 \begin{tabular}{c c}
 \hline Parameter & Value  \\
 \hline
 $V_0$ & 0.04 \Tstrut\\
 $\kappa$ & 2 \\
 $m$ & 0.04 \\
 $\xi$ & 0.25\\
 $\rho$ & -0.5\\
 \hline
 \end{tabular}
\caption{SV parameters for the synthetic data example} \label{tab:syn_sv_parameters}
\end{center}
\end{table}

\subsection{Real Data}\label{sec:real}

\begin{figure}[H]
\centering
\includegraphics[width=0.5\textwidth]{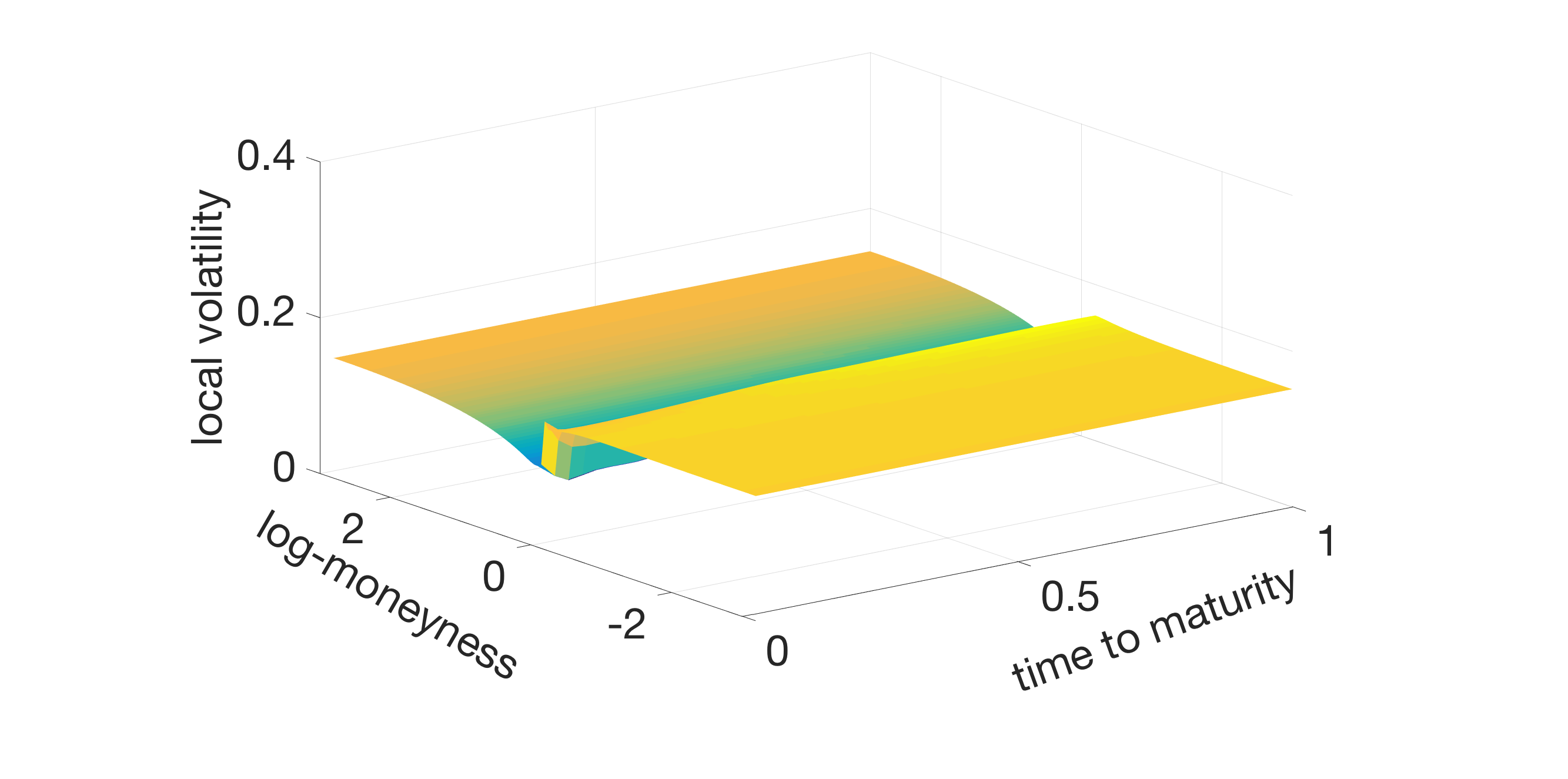}\hfill
\includegraphics[width=0.5\textwidth]{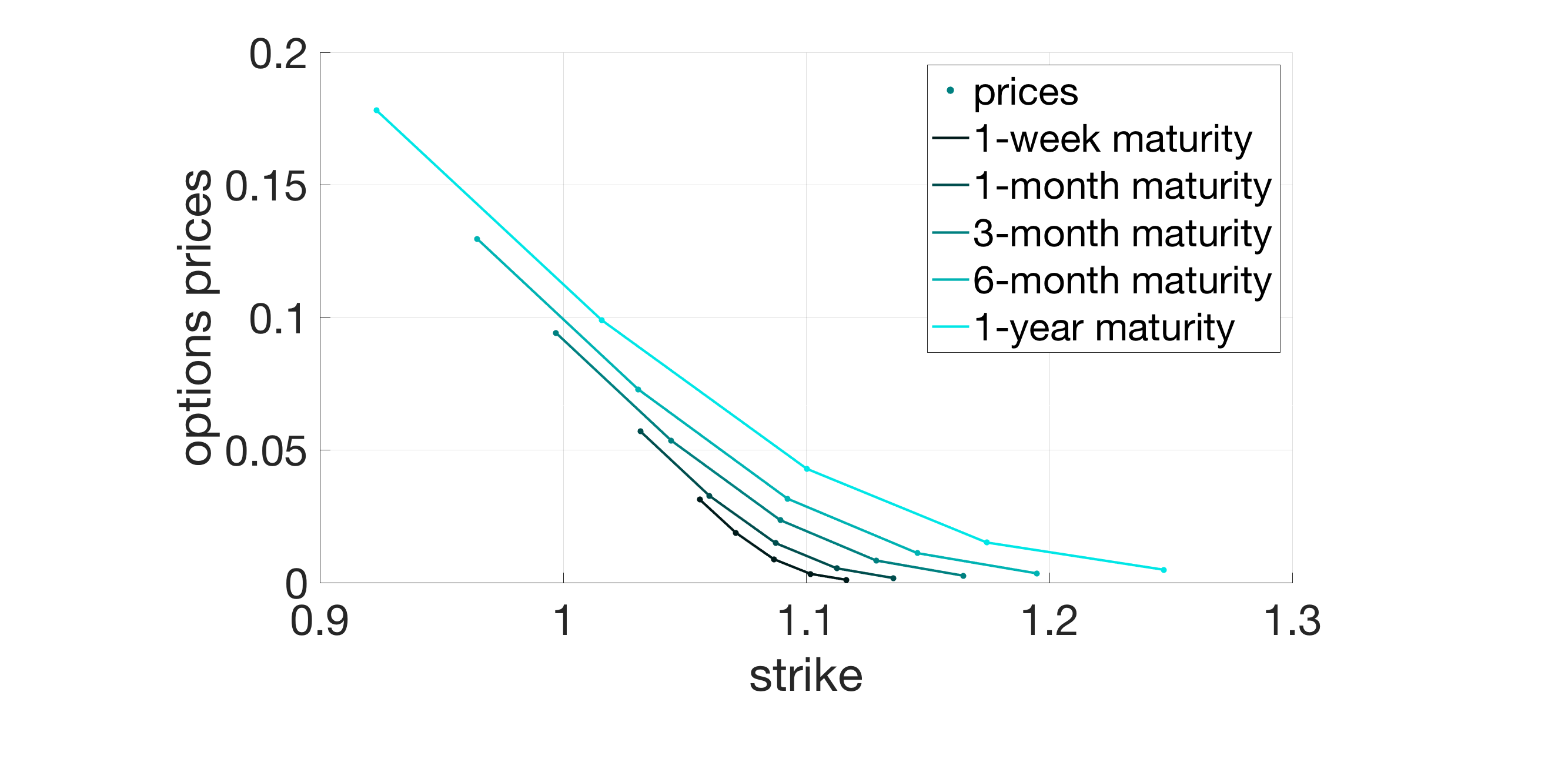}\hfill
\caption{EUR-USD local volatility surface and options prices on March 18th, 2015}\label{real1}
\end{figure}

In this section we present a real data example. We chose FX options on EURUSD on March 18th, 2015. They include the typical 25 liquid option contracts, with 5 maturities (1W, 1M, 3M, 6M, 1Y) and 5 strikes (related to 10 and 25 Call and Put Delta and to ATM) per maturity (see Figure \ref{real1}). The spot value was 1.0864. The parameters of the Heston model are calibrated to this data set and given in Table \ref{tab:real_sv_parameters}.
\begin{table}[h!]
\begin{center}
 \begin{tabular}{c c}
 \hline 
 Parameter & Value  \\
 \hline 
 $V_0$ & 0.013 \Tstrut\\
 $\kappa$ & 1.025 \\
 $m$ & 0.013 \\
 $\xi$ & 0.161\\
 $\rho$ & -0.626\\
 \hline
 \end{tabular}
\caption{SV parameters calibrated to real data} \label{tab:real_sv_parameters}
 \end{center}
\end{table}

\begin{figure}[H]
\centering
\includegraphics[width=0.5\textwidth]{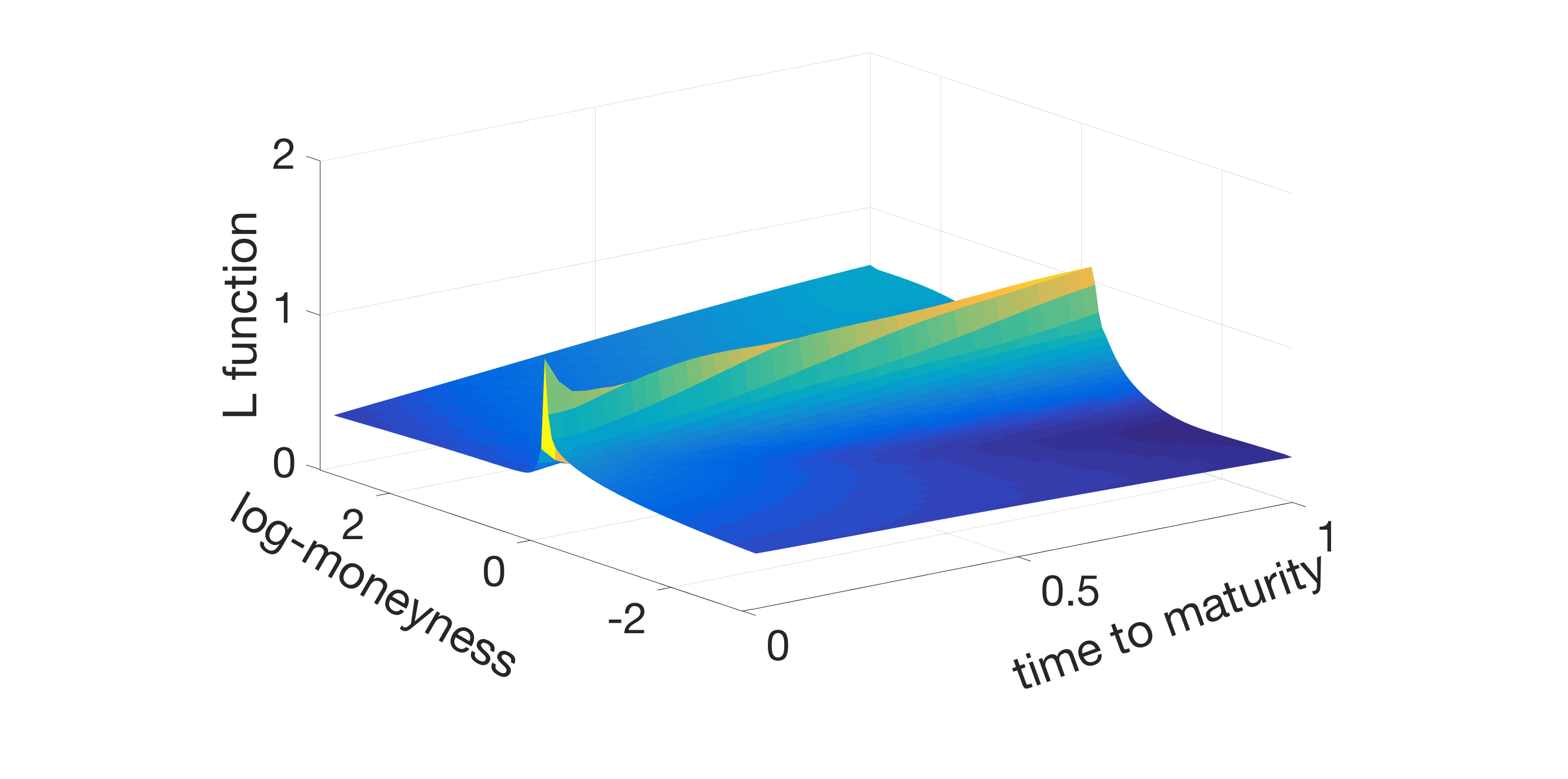}\hfill
\includegraphics[width=0.5\textwidth]{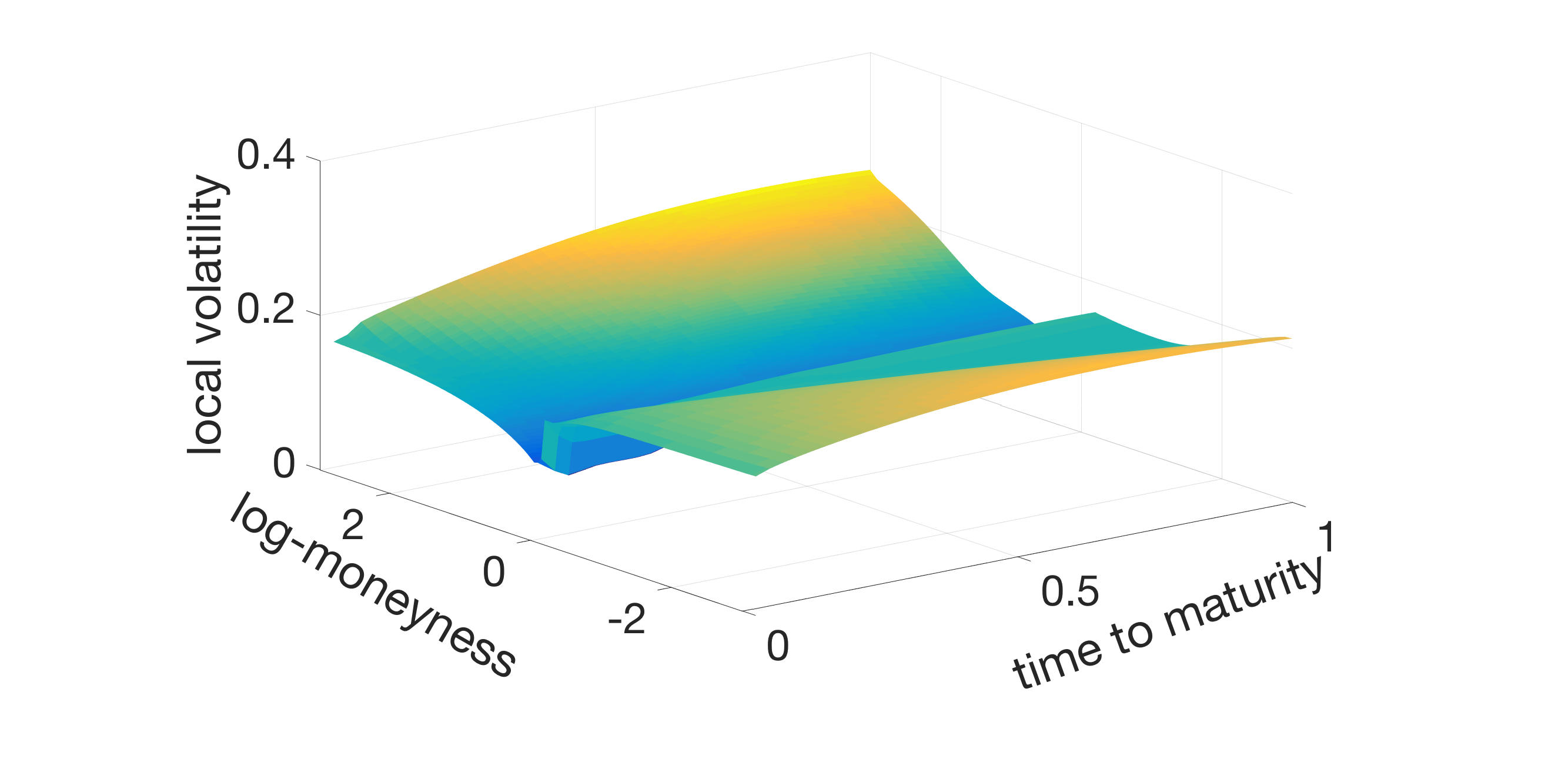}\hfill
\caption{Leverage function (left) and the local volatility surface (right) computed with the benchmark method in the real data example.}\label{real2}
\end{figure}

\begin{figure}[H]
\centering
\includegraphics[width=0.5\textwidth]{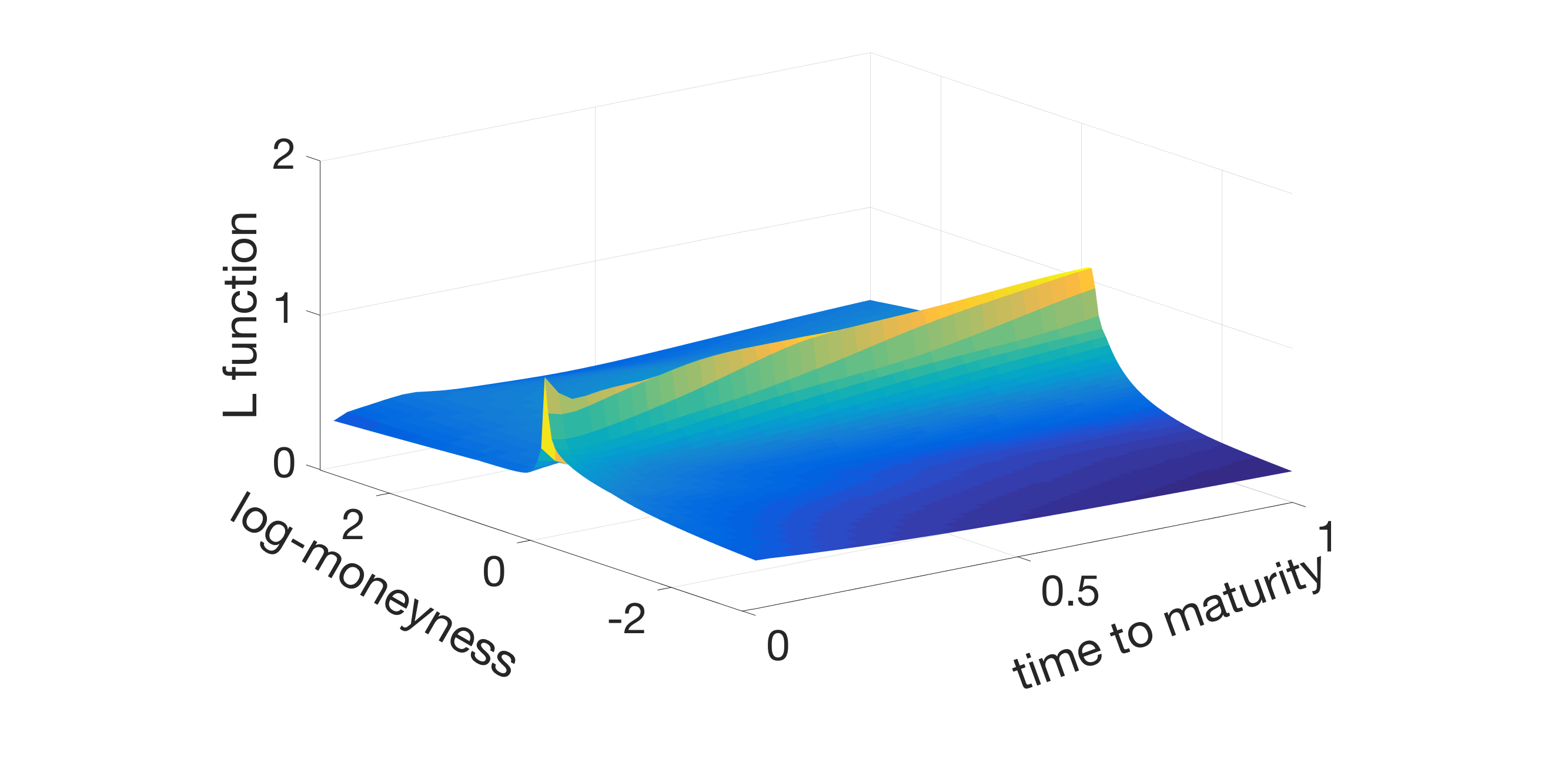}\hfill
\includegraphics[width=0.5\textwidth]{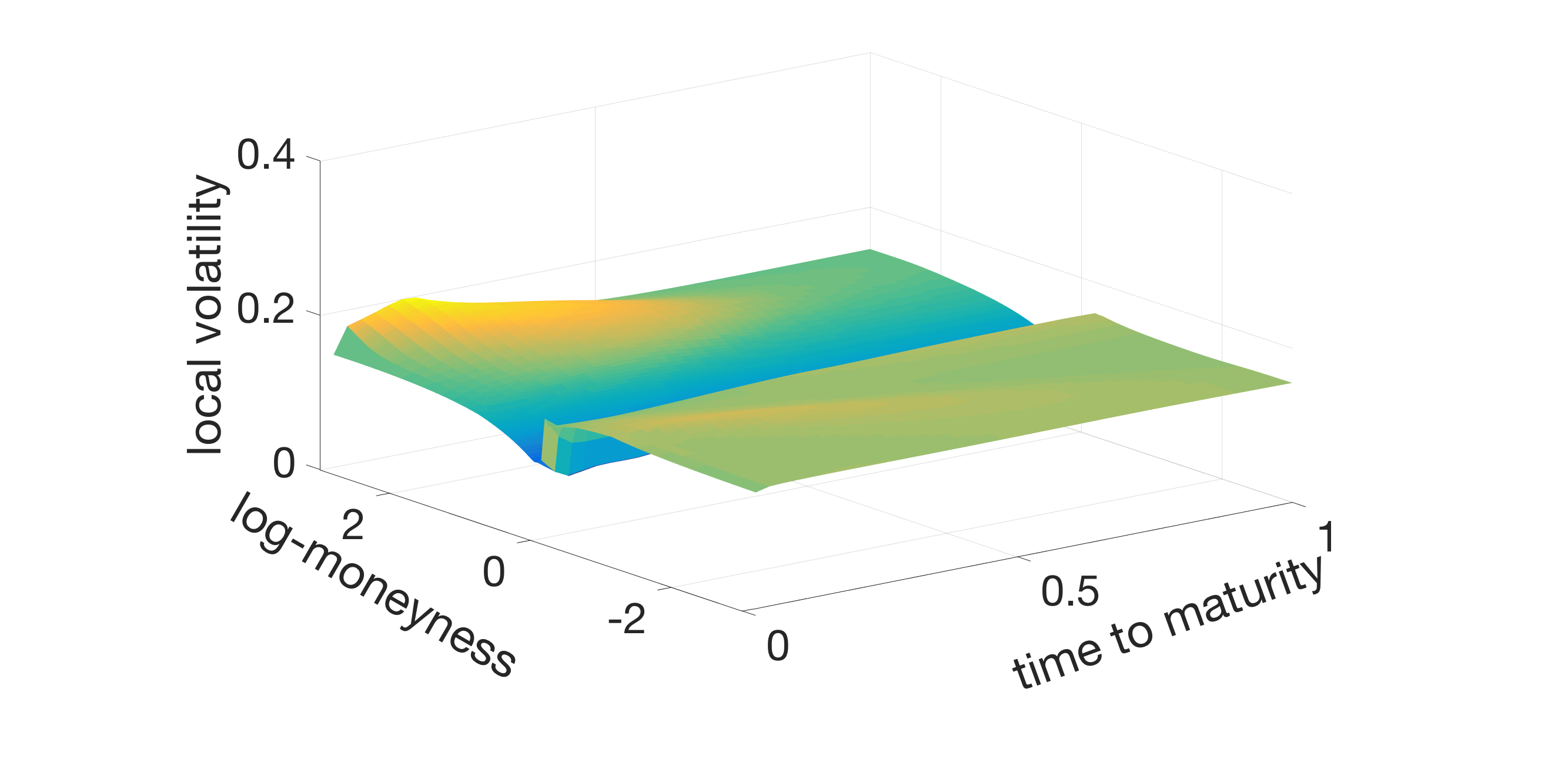}\hfill
\caption{Leverage function (left) and the local volatility surface (right) computed with our proposed method in the real data example.}\label{real3}
\end{figure}

These parameters were required to satisfy the Feller condition. This translates into more realistic dynamics for the volatility, since it prevents the volatility process $V$ to reach the zero boundary. The domestic and foreign interest rates are the same as in Figure \ref{fig:interest_rates}. We choose the same discretization parameters as in the synthetic data example, see Table \ref{tab:mesh_parameters}. In Figure \ref{real1}, we show the estimated local volatility surface from option prices. For the description of the methods that we used to calibrate the local volatility surface, see \cite{albani2015}. In Figure \ref{real2}, we show the recovered local volatility surface and the leverage function. In Figure \ref{real3}, we implemented the benchmark method. The Tikhonov parameters are $\alpha_1=0$ and $\alpha_2=10^{-3}$.

\subsection{Numerical Results}

In the figures below we show the recovered leverage function and the local volatility surface using the benchmark method in Section \ref{sec:benchmark_method} and our proposed method shown in Section \ref{sec:proposed_method}.

\subsubsection*{Synthetic Data}

\begin{figure}[H]
\centering
\includegraphics[width=0.5\textwidth]{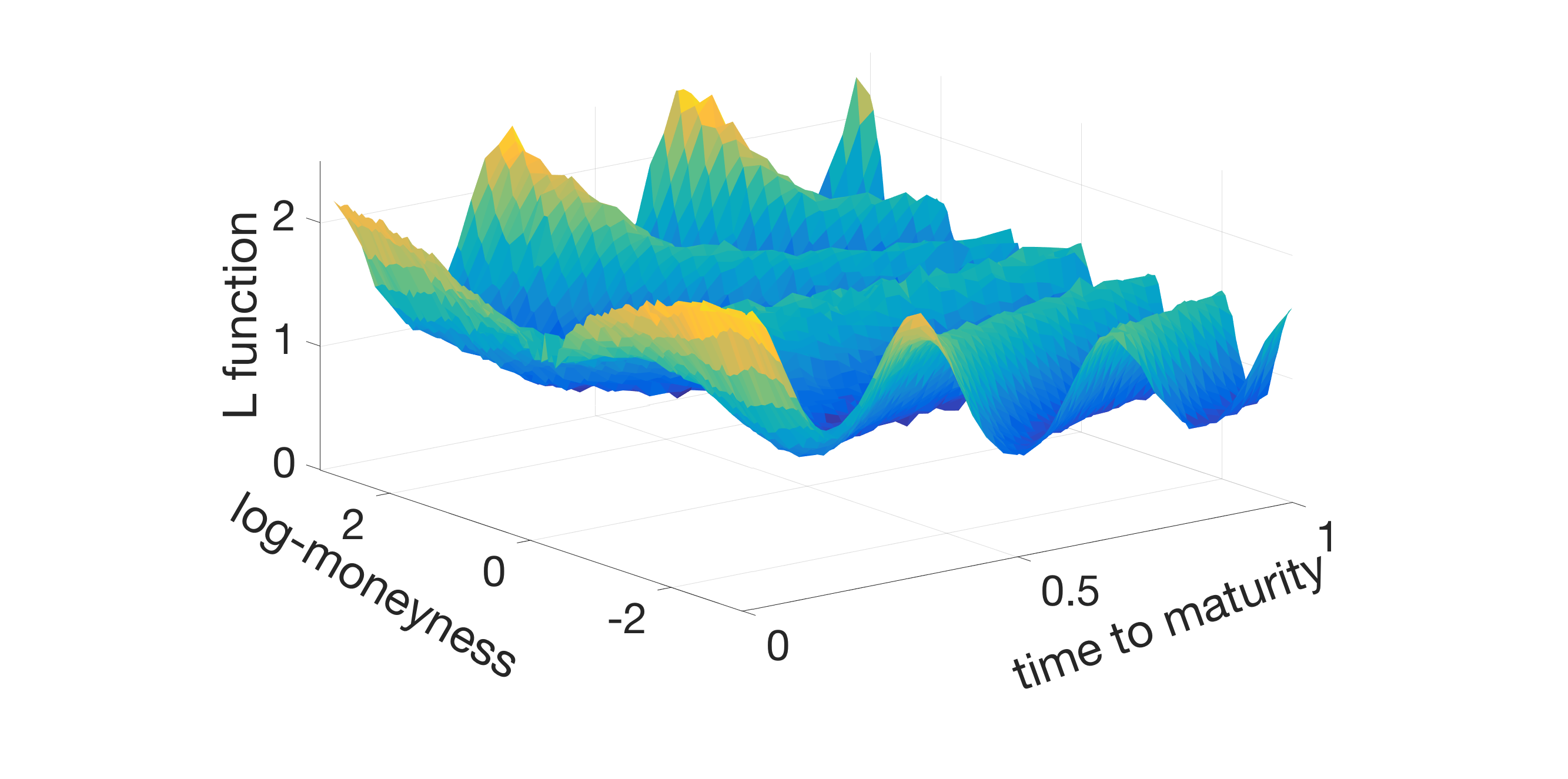}\hfill
\includegraphics[width=0.5\textwidth]{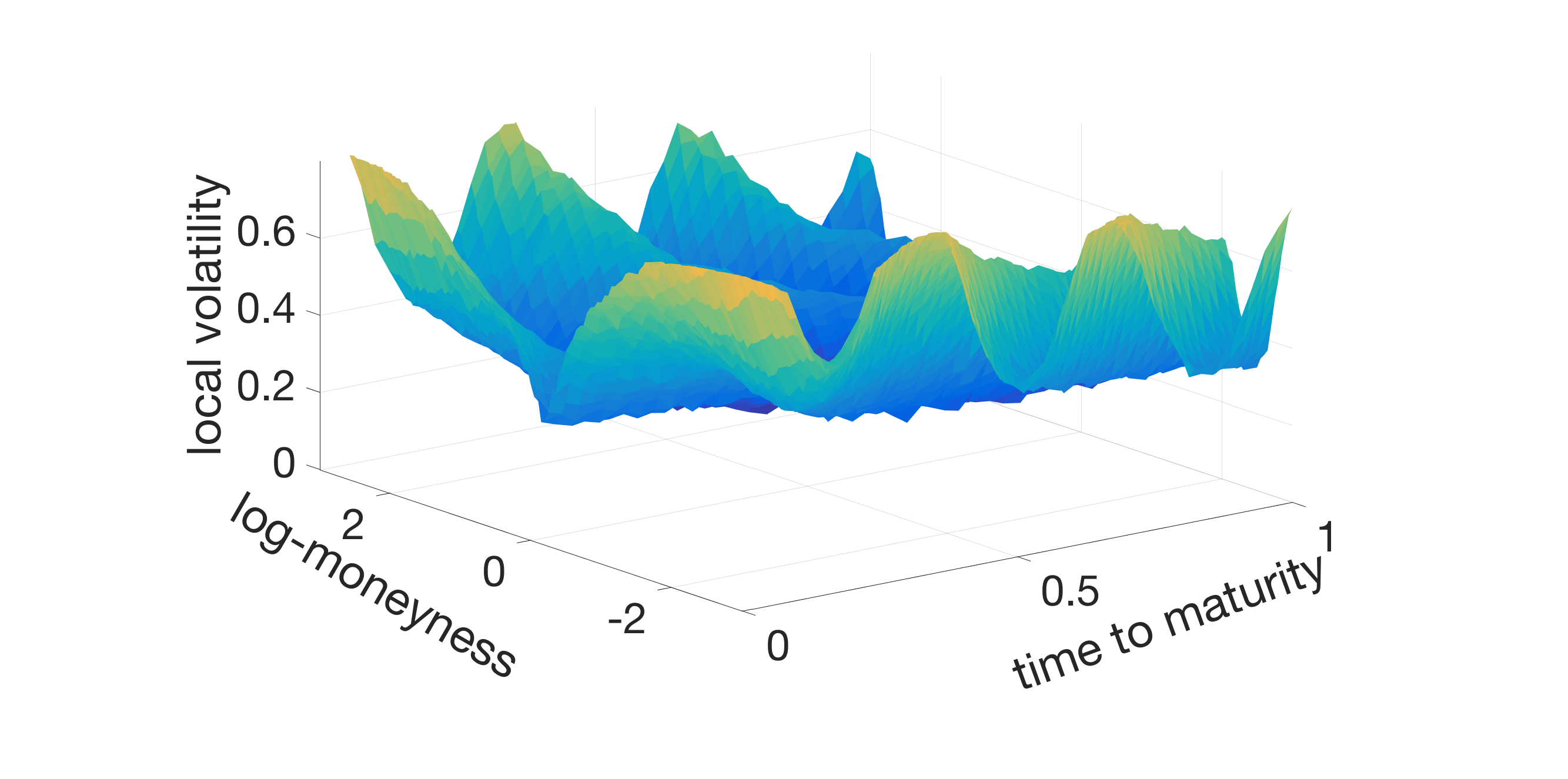}\hfill
\caption{Leverage function (left) and the local volatility surface (right) computed with the benchmark method in the synthetic data example.}\label{syn2}
\end{figure}

\begin{figure}[H]
\centering
\includegraphics[width=0.5\textwidth]{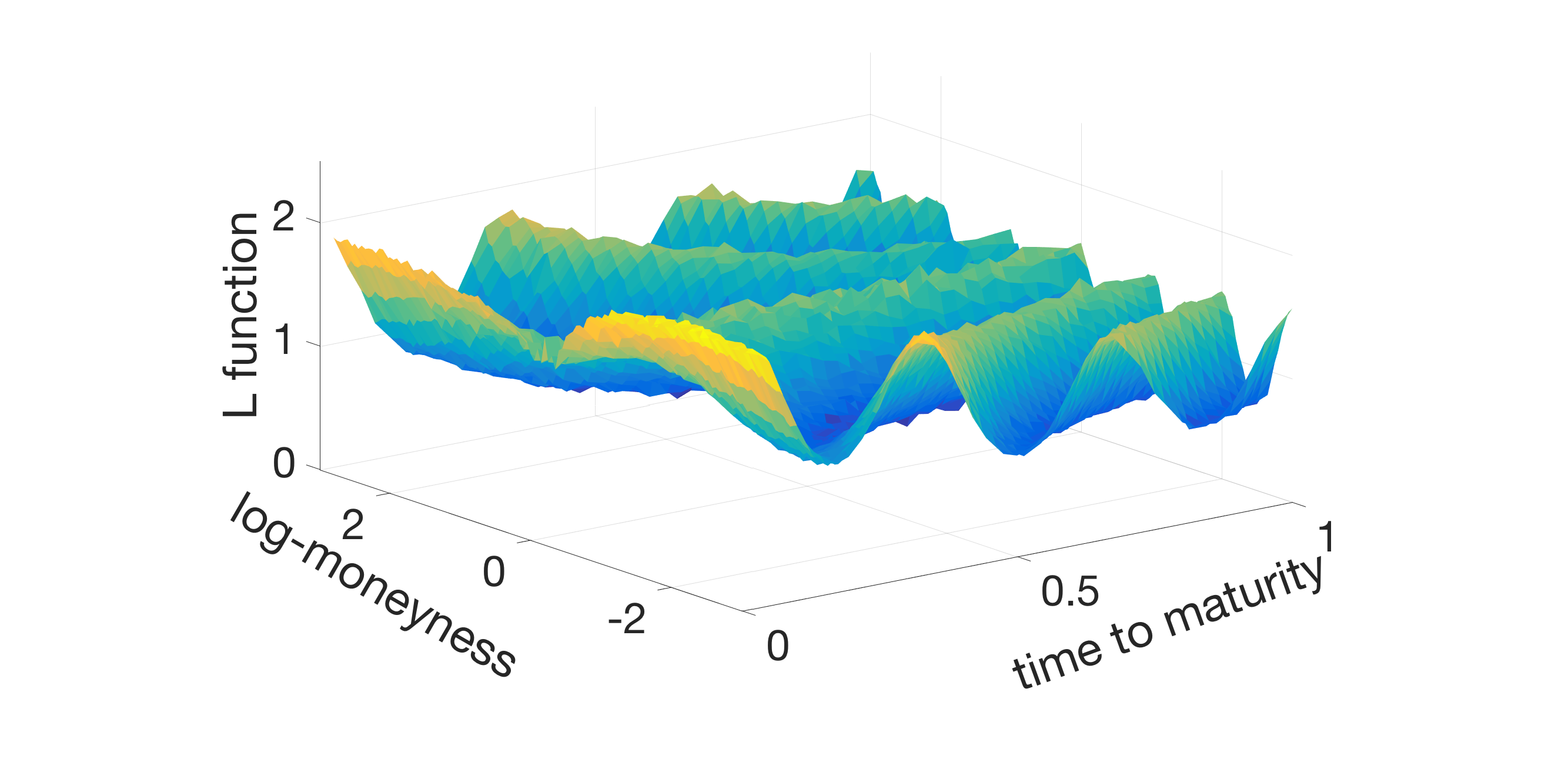}\hfill
\includegraphics[width=0.5\textwidth]{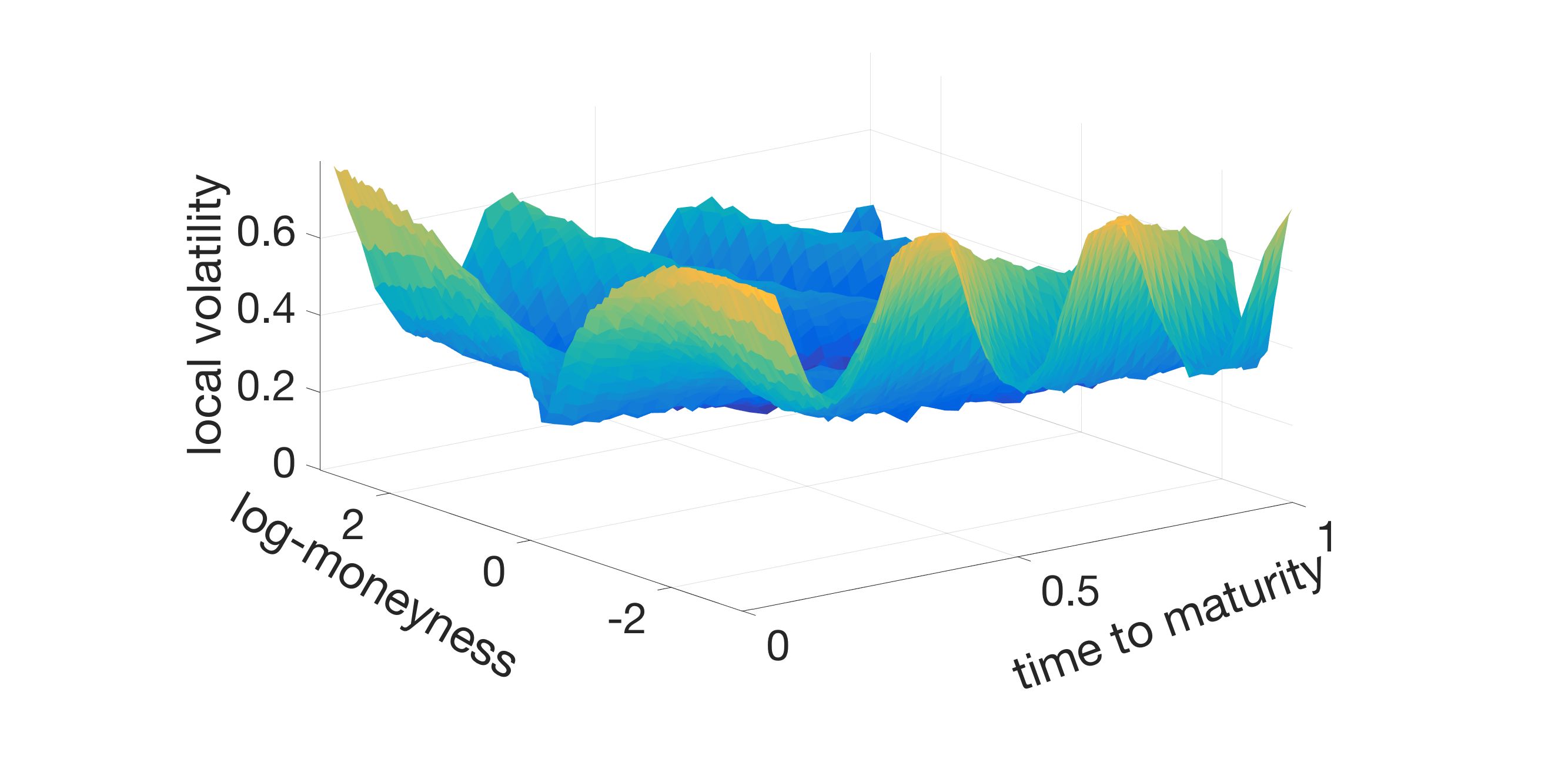}\hfill
\caption{Leverage function (left) and the local volatility surface (right) computed with our proposed method in the synthetic data example.}\label{syn3}
\end{figure}

\begin{figure}[H]
\centering
\includegraphics[width=0.33\textwidth]{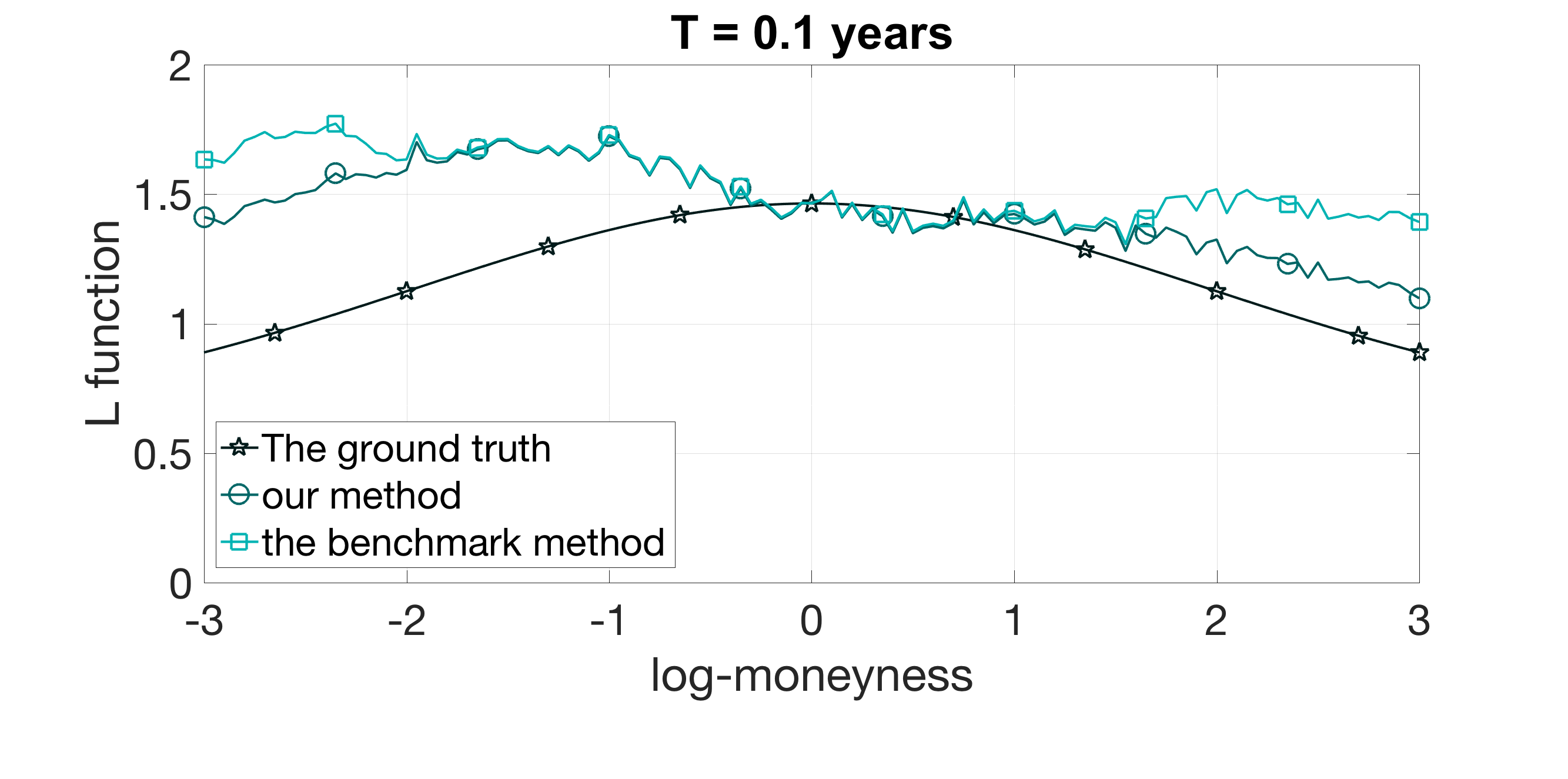}\hfill
\includegraphics[width=0.33\textwidth]{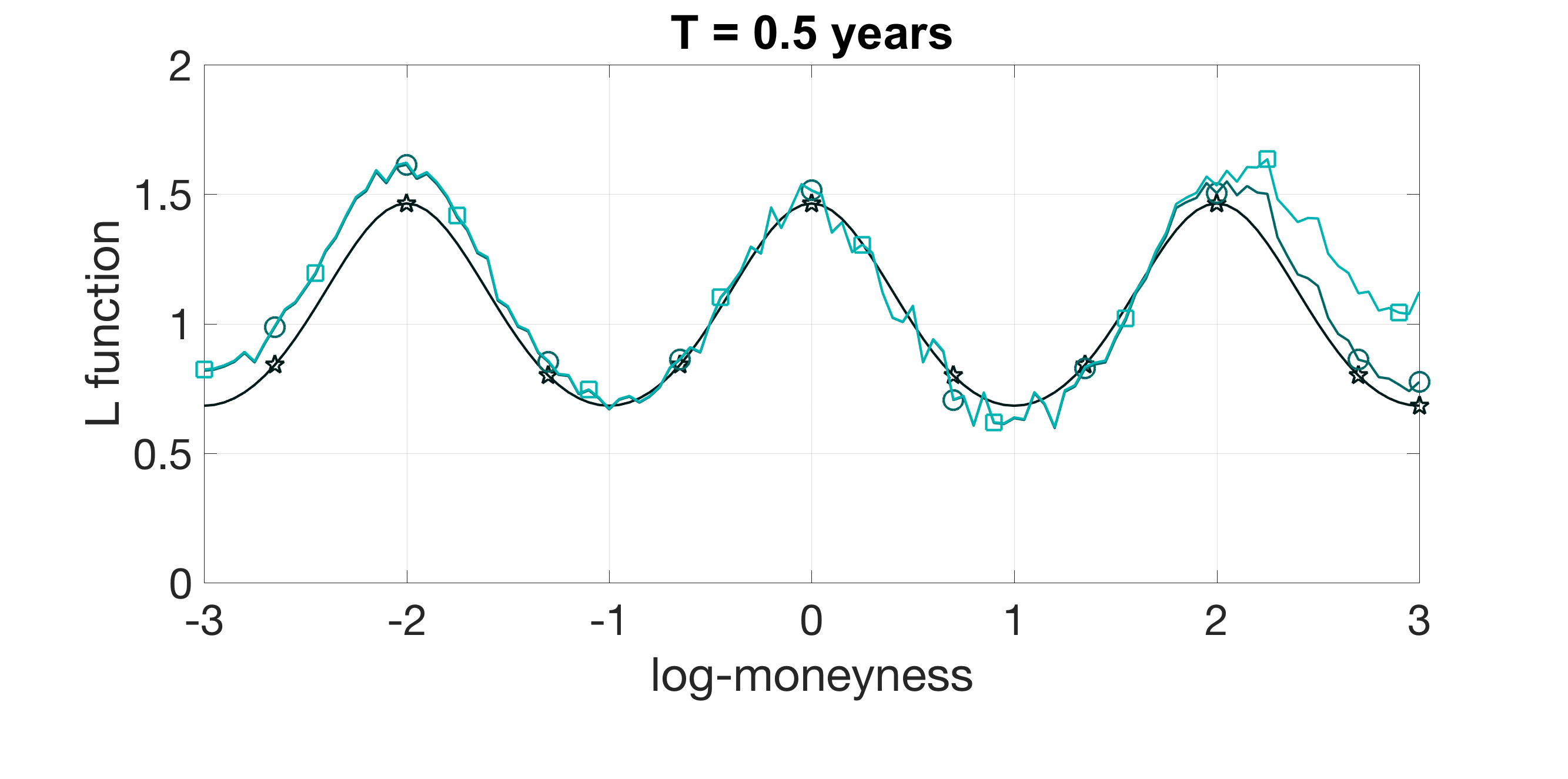}\hfill
\includegraphics[width=0.33\textwidth]{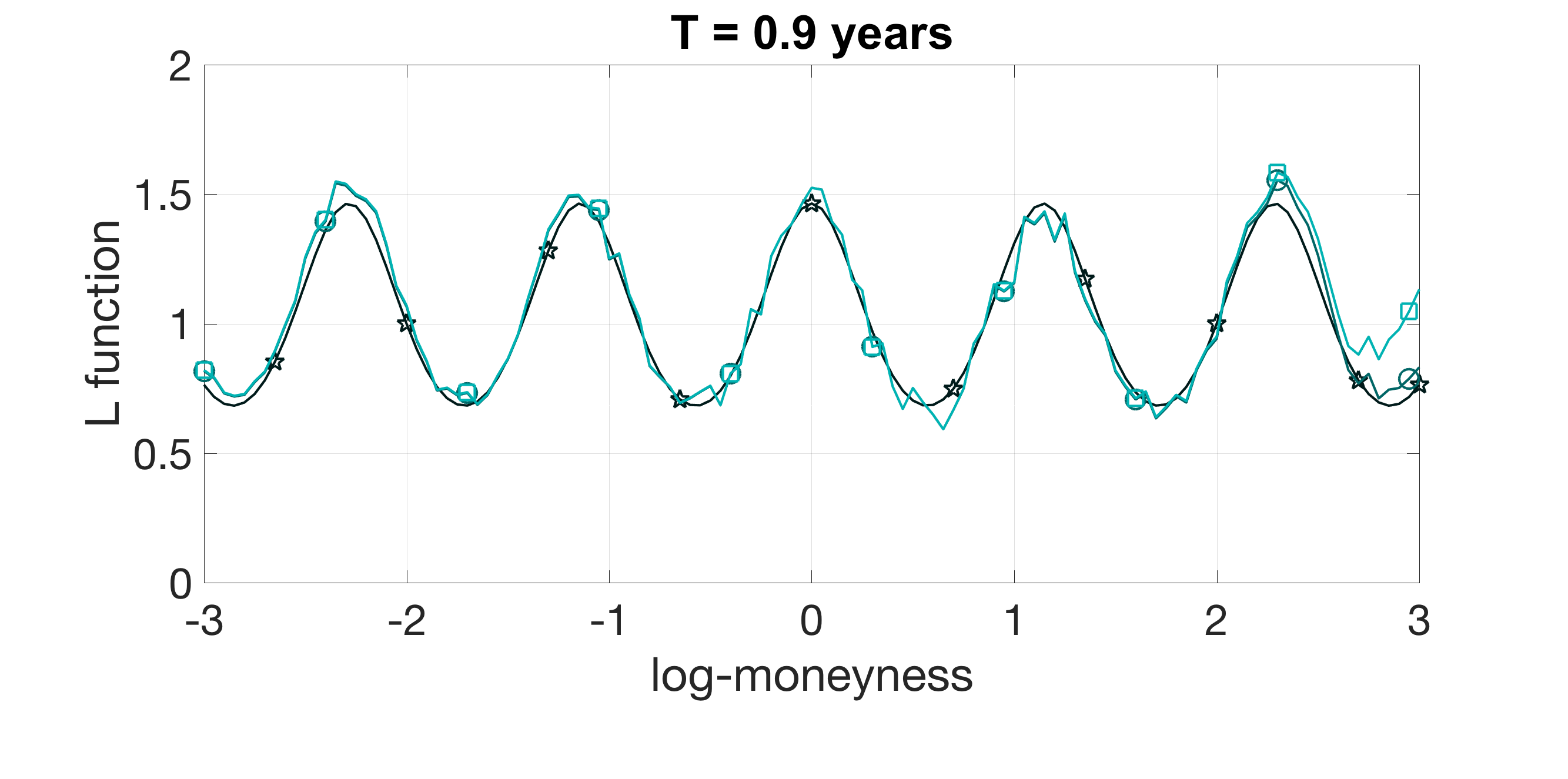}\hfill
\caption{The leverage function in the synthetic data example:  the ground truth (with stars), the benchmark method (with squares) and our method (with circles)}\label{syn4}
\end{figure}

\begin{figure}[H]
\centering
\includegraphics[width=0.33\textwidth]{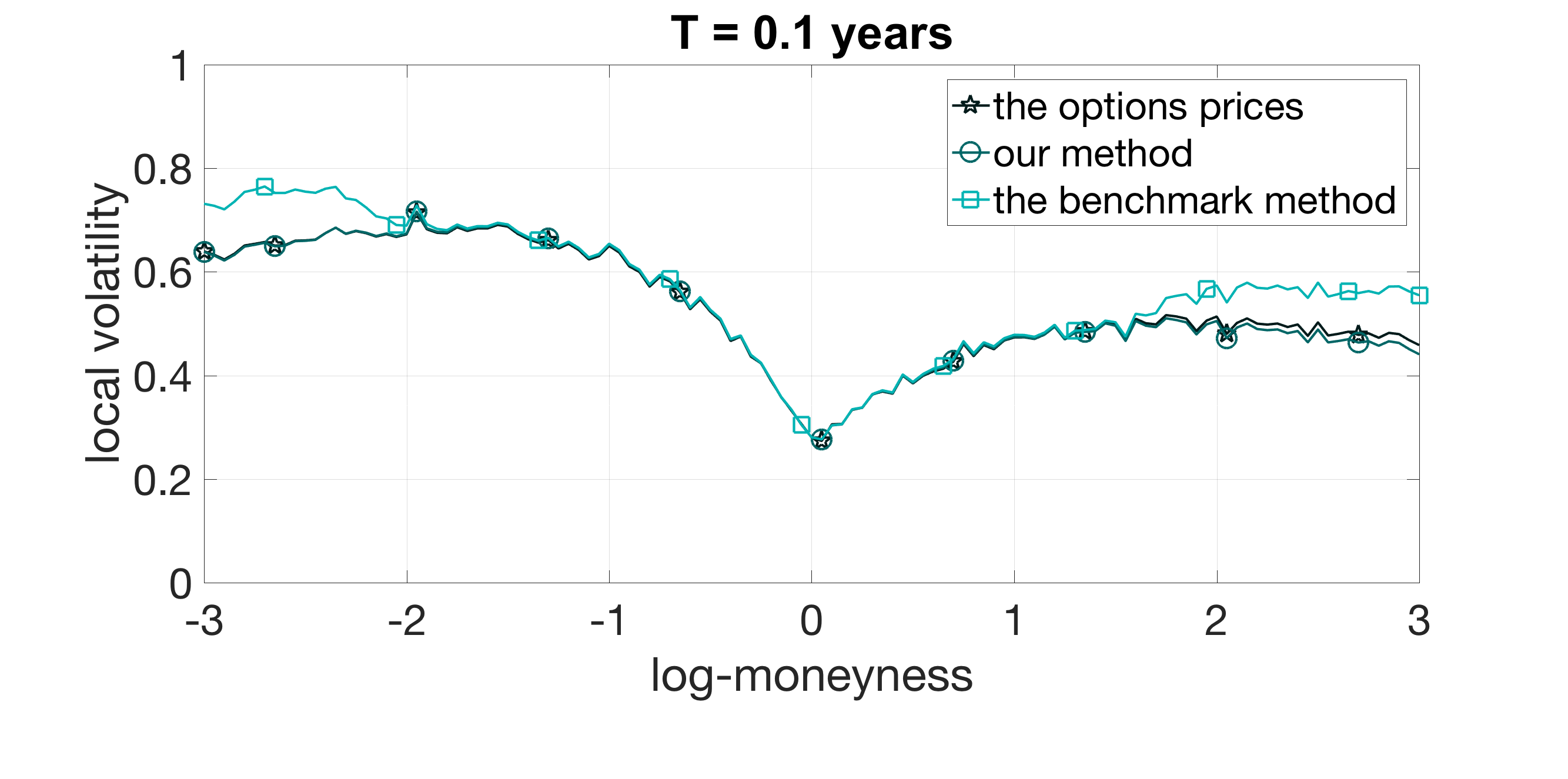}\hfill
\includegraphics[width=0.33\textwidth]{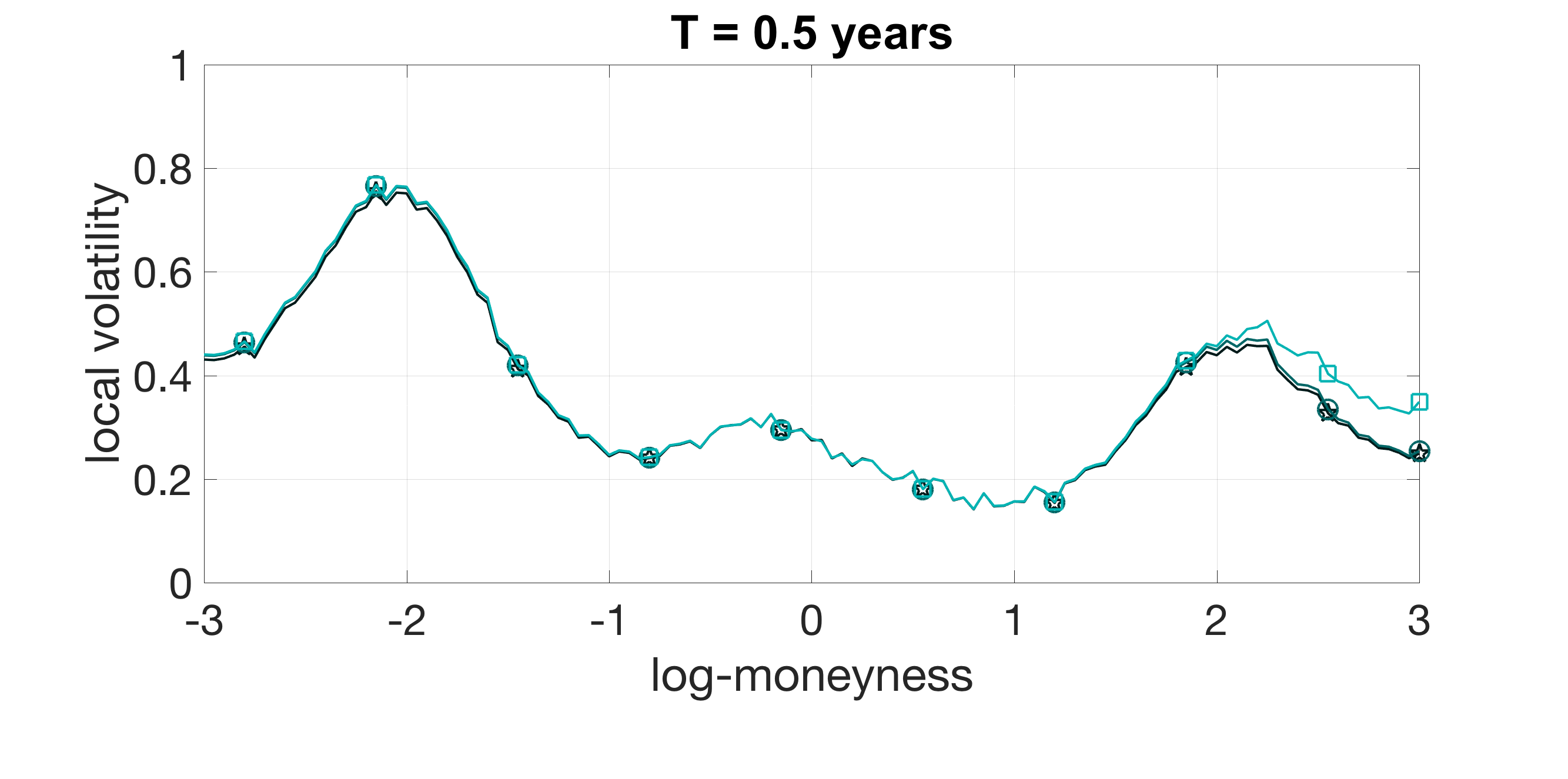}\hfill
\includegraphics[width=0.33\textwidth]{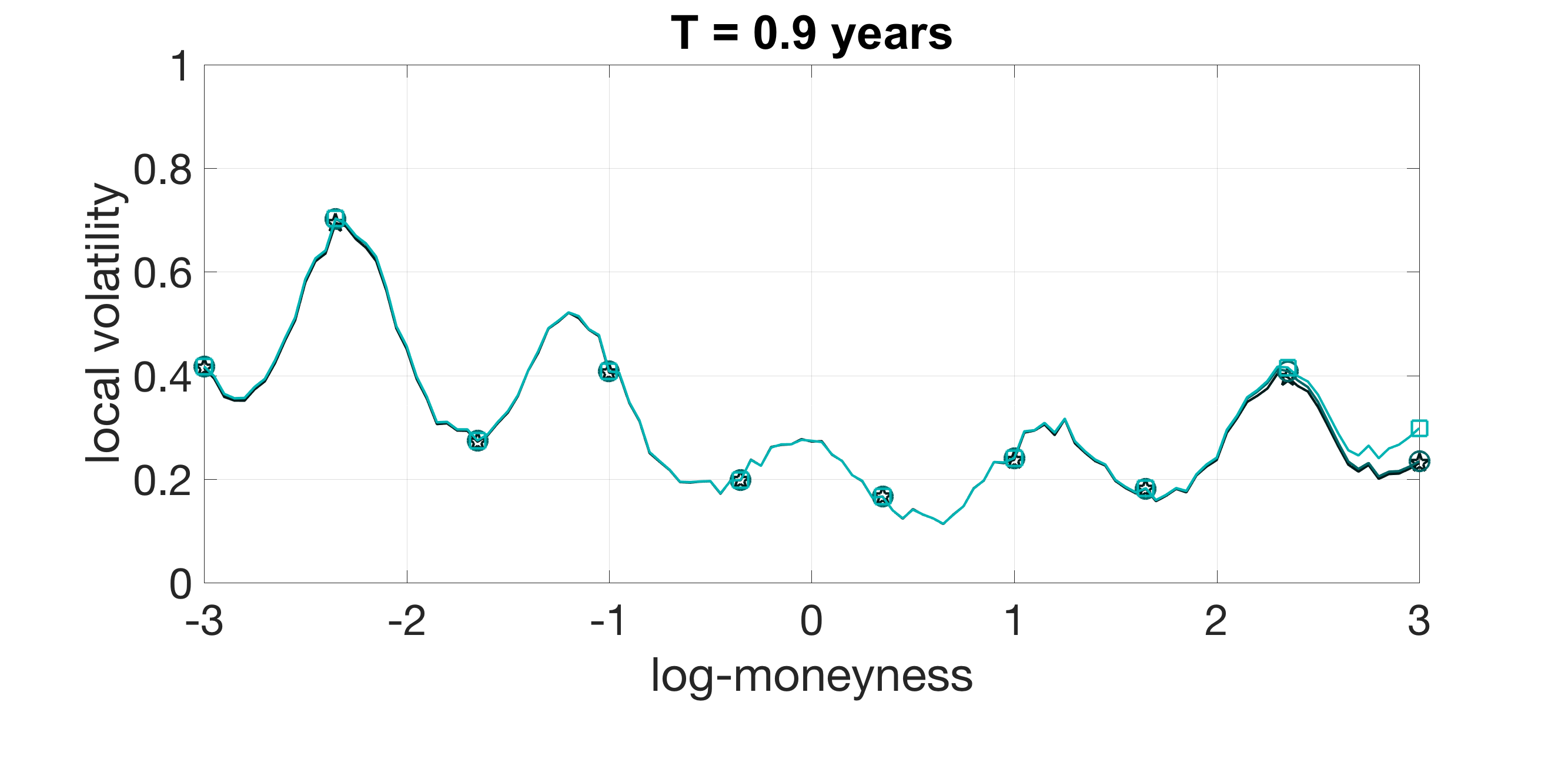}\hfill
\caption{The local volatility surface in the synthetic data example:  the ground truth (with stars), the benchmark method (with squares) and our method (with circles)}\label{syn5}
\end{figure}

\subsubsection*{Real Data}

\begin{figure}[H]
\centering
\includegraphics[width=0.33\textwidth]{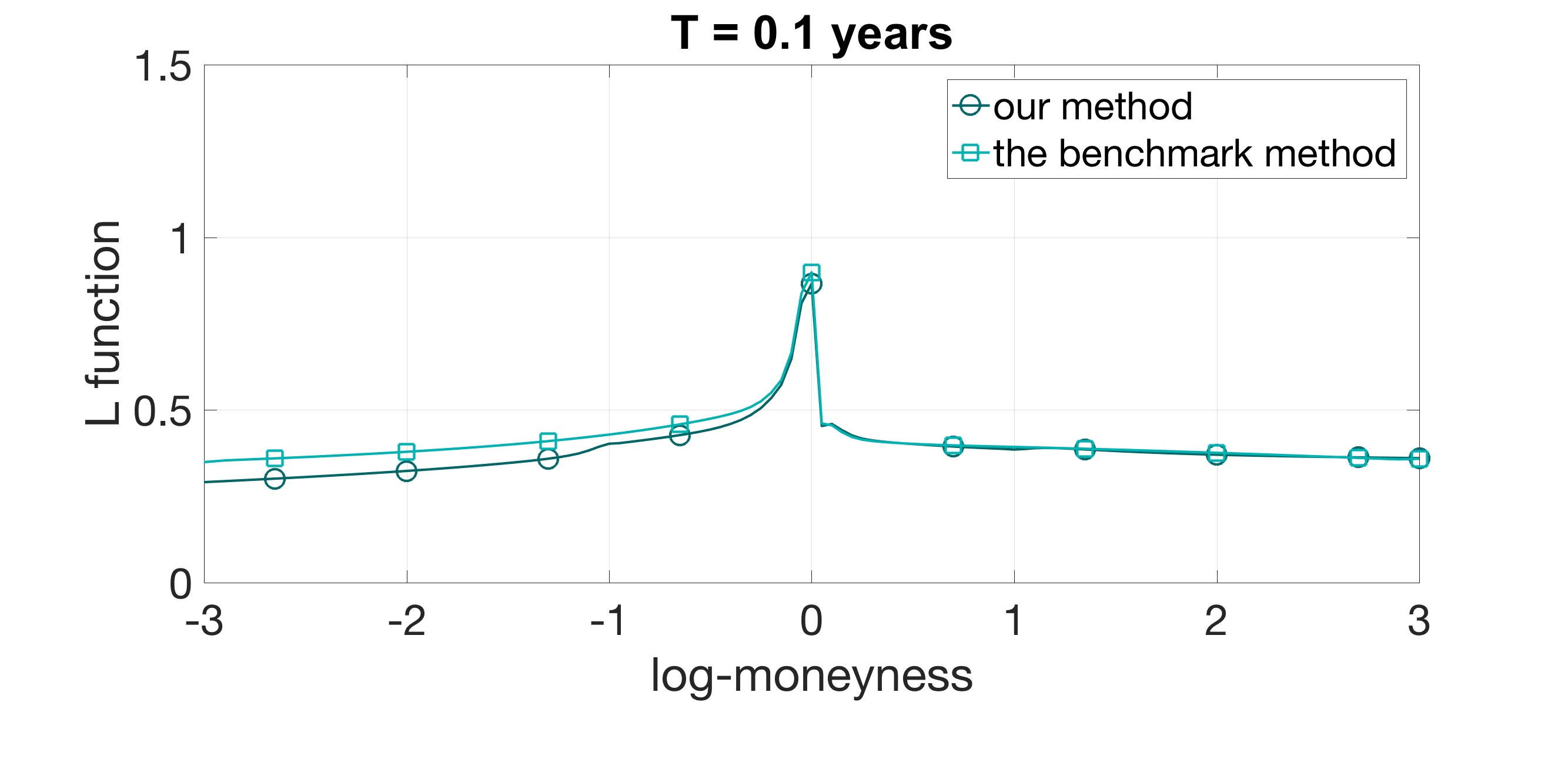}\hfill
\includegraphics[width=0.33\textwidth]{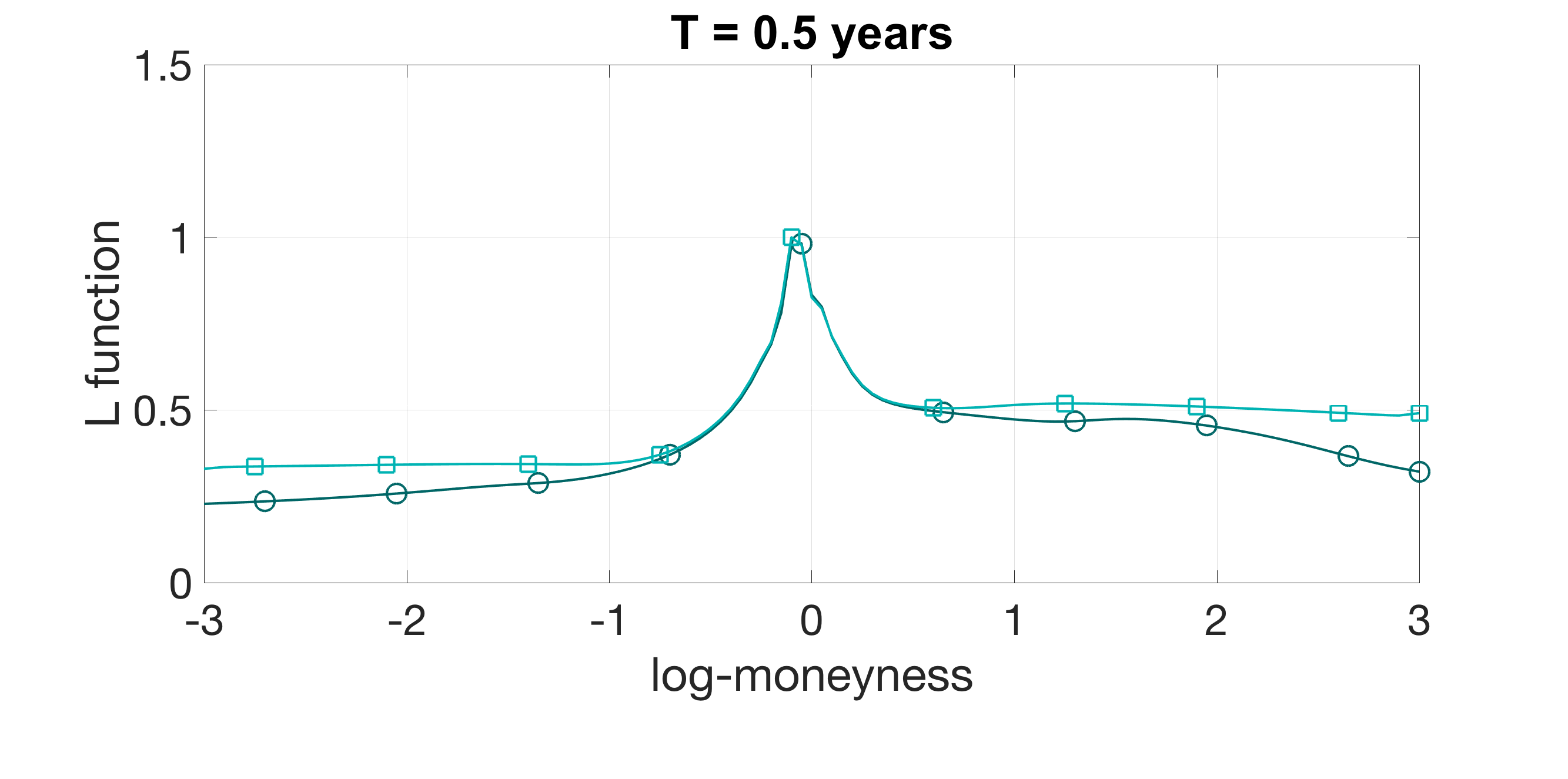}\hfill
\includegraphics[width=0.33\textwidth]{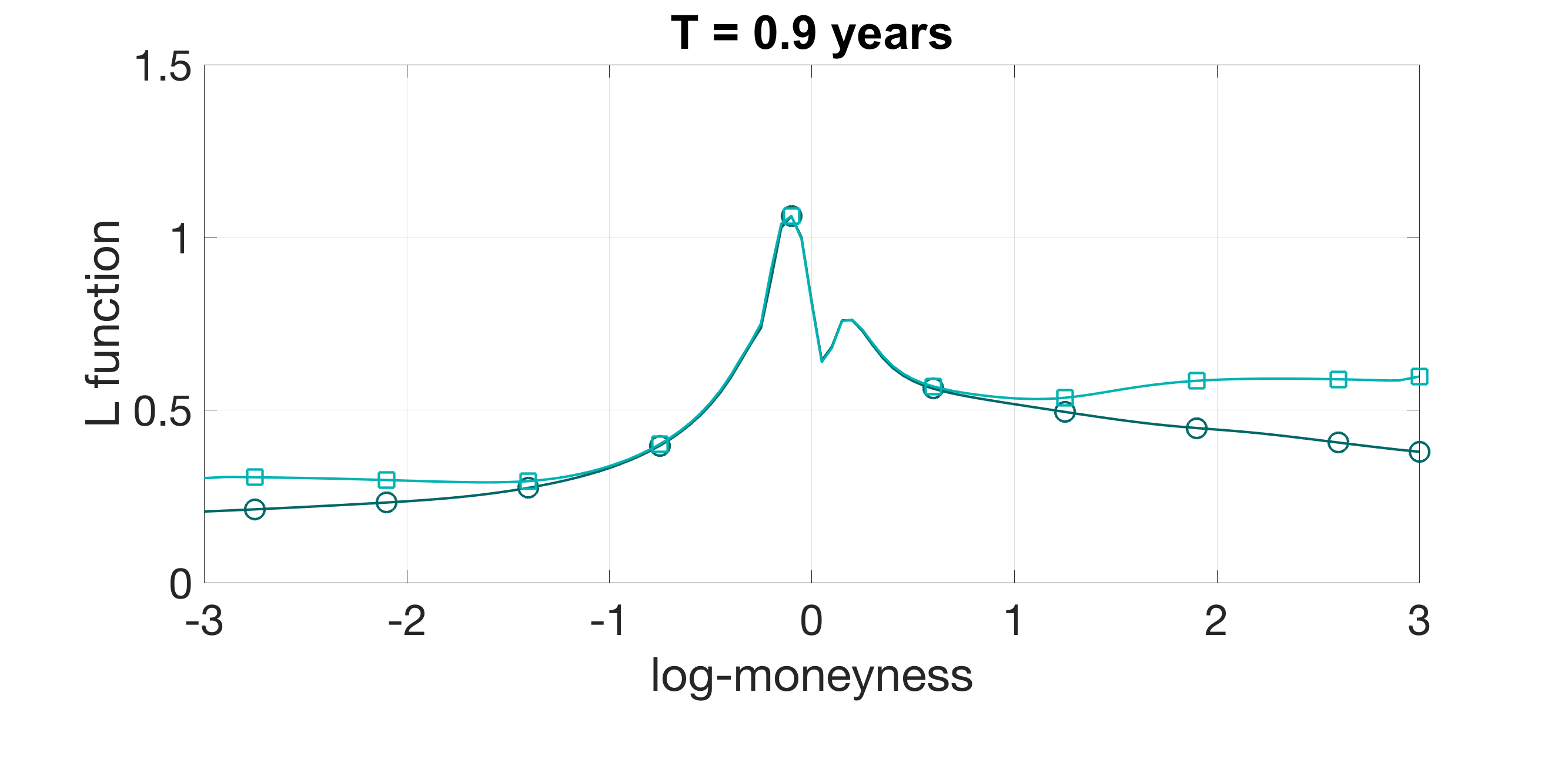}\hfill
\caption{The leverage function in the real data example:  the benchmark method (with squares) and our method (with circles)}\label{real4}
\end{figure}

\begin{figure}[H]
\centering
\includegraphics[width=0.33\textwidth]{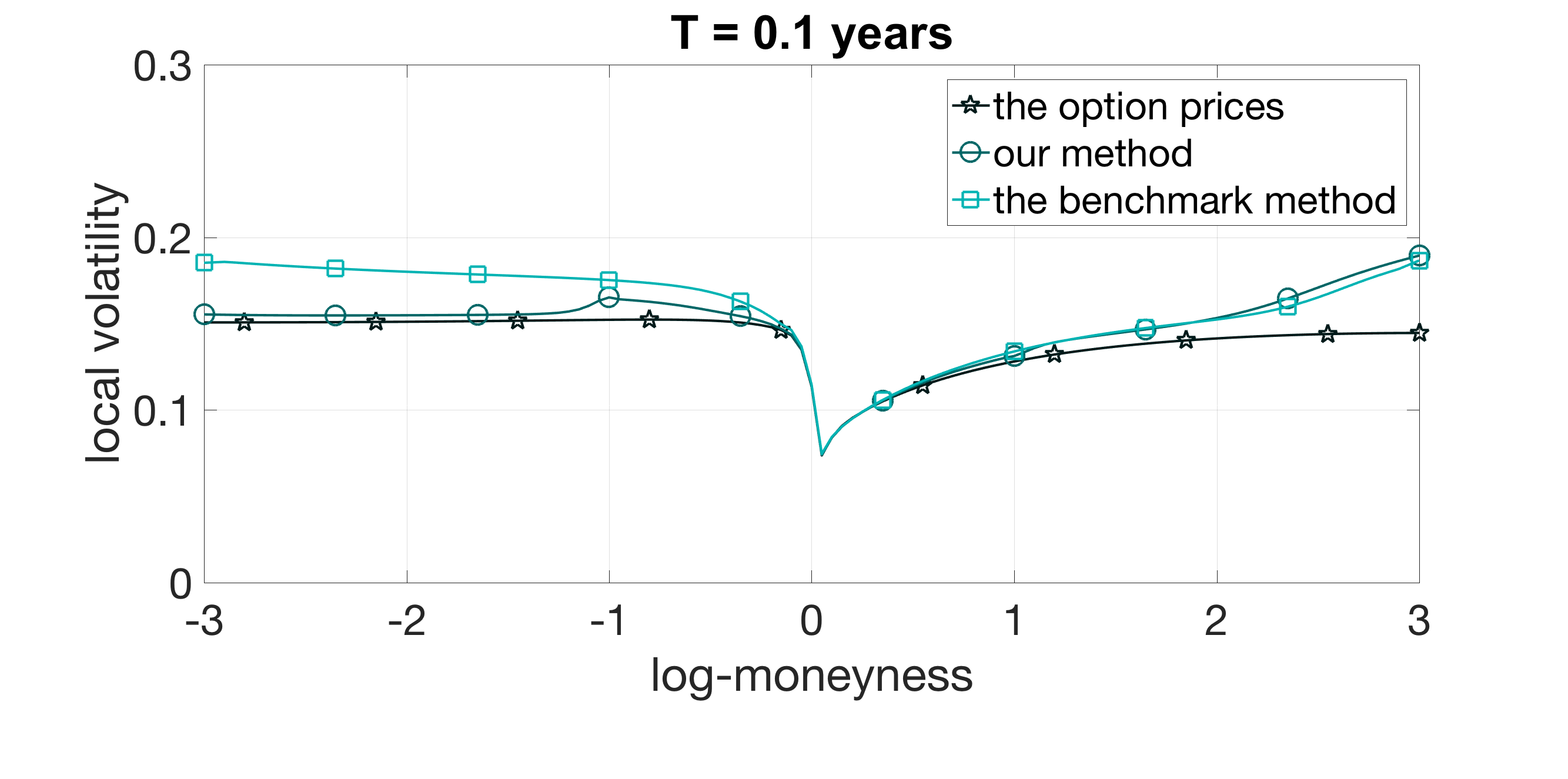}\hfill
\includegraphics[width=0.33\textwidth]{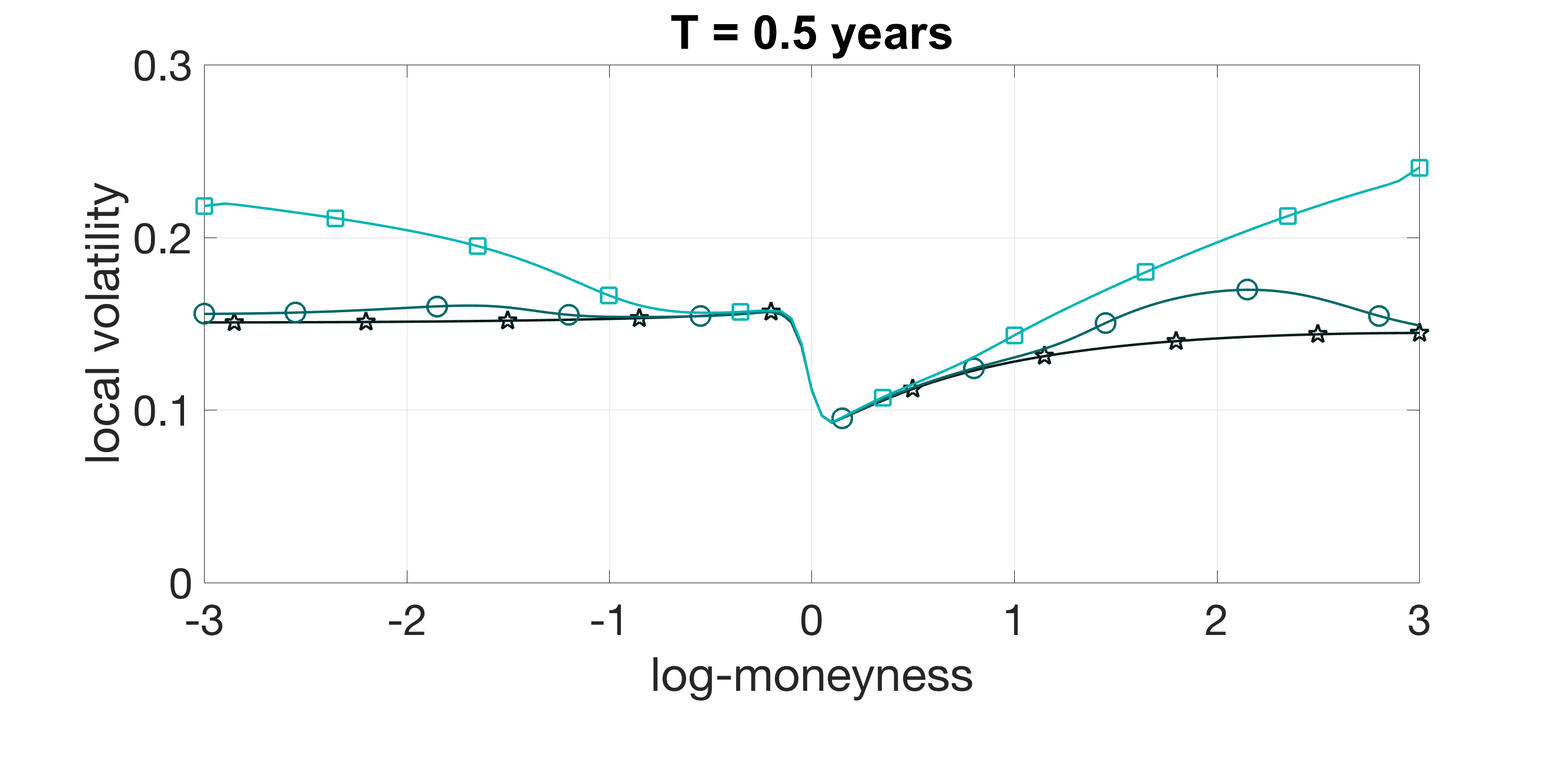}\hfill
\includegraphics[width=0.33\textwidth]{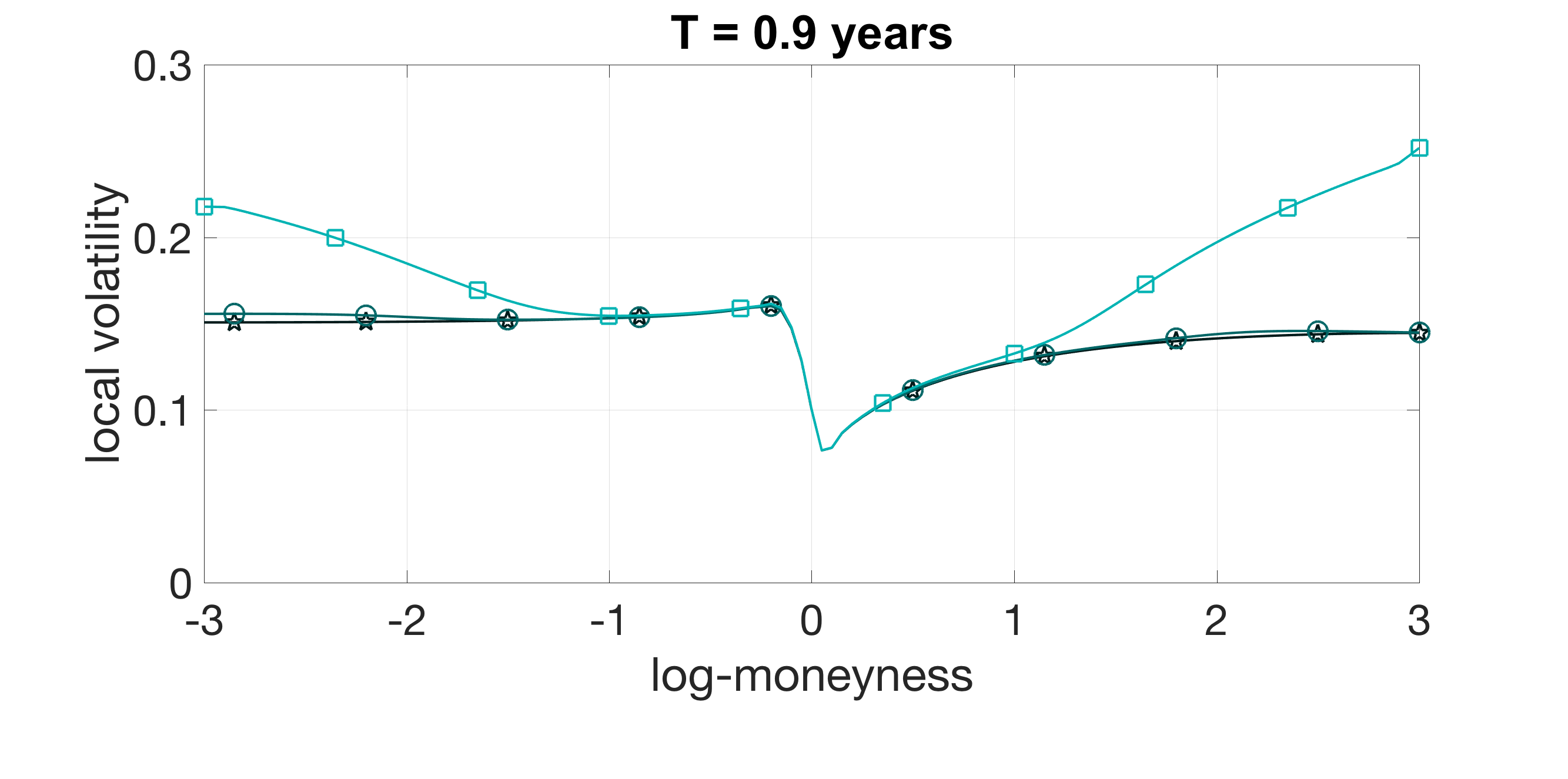}\hfill
\caption{The local volatility surface in the real data example: computed from option prices (with stars), the benchmark method (with squares) and our method (with circles)}\label{real5}
\end{figure}

\subsection{Conclusions}\label{sec:conclusion}

From the figures shown in the previous subsection, it can be seen that 

\begin{itemize}

\item[$\diamond$] the benchmark method is not stable and it fails to converge for large log-moneyness;

\item[$\diamond$] the results of the benchmark method have more noise;

\item[$\diamond$] for larger maturities the proposed method converges to the ground truth leverage function.

\end{itemize}

In Figures \ref{syn4}, \ref{syn5}, \ref{real4} and \ref{real5}, we show the recovered leverage function and local volatility for the two methods at 3 different times. The proposed method and the benchmark method agree around at-the-money, but for deep in-the-money and out-of-the money log-moneyness, the corresponding local volatility curve of the benchmark method is distant from the market's local volatility surface. We have shown three maturities, but this phenomenon can also be observed for the other maturities. This phenomenon is observed less prominently in the synthetic data example and the reason is that the local volatility surface is smoother.

Once we estimate the leverage function $L$, we can recover the local volatility surface using the Alternating Direction Implicit (ADI) method for the Fokker-Planck PDE with the leverage function at both times, $t_n$ and $t_{n+1}$, see Section \ref{sec:numerical_fp}. Comparing with the ground truth of local volatility surface in the synthetic data example, we can calculate the relative residuals. In Table \ref{tab:residual}, we present the relative residuals in two intervals of the log-moneyness, which are $[-3,3]$ and $[-2,2]$. We also report the relative residuals of the real data example.  For both examples, we see that the proposed method generates better results with relative errors significantly smaller than the benchmark method. We would like to point out that this failure of convergence of the benchmark method is not related to a boundary issue. Indeed, numerical experiments on smaller log-moneyness intervals have similar results to the truncated version of the results we have found.

\begin{table}[h!]
\begin{center}
 \begin{tabular}{l | c c }
\hline 
Example & Benchmark in $[-3,3]$ (in $[-2,2]$) & Proposed in $[-3,3]$ (in $[-2,2]$)  \\
\hline
Synthetic & 7.92\% (2.07\%) & 1.40\% (1.09\%) \Tstrut\\
Real & 27.82\% (15.44\%) & 7.93\% (5.44\%)\\
 \hline
 \end{tabular}
\caption{Relative residuals} \label{tab:residual}
\end{center}
\end{table}

The conclusion from our numerical exercises, that corroborates the theoretical reasoning, is that, when compared to the benchmark, the proposed method
\begin{itemize}
 \item[$\diamond$] is more robust against noise;
 \item[$\diamond$] is more resilient to instabilities in the regions of low probability density of the spot prices and instantaneous variance;
 \item[$\diamond$] does not require \textit{ad hoc} procedures to avoid instabilities due to low probability regions.
\item[$\diamond$] respects the data in the sense that we do not apply interpolation. More precisely, the benchmark method requires the knowledge of the local vol on the same mesh as the one used for the Fokker-Planck Equation (\ref{eq:fokker_planck}).
\end{itemize}

\section{Concluding Remarks}
\label{sec:concluding_remarks}

We have studied the calibration of the Stochastic-Local Volatility model and proposed a numerical method based on the Tikhonov regularization framework. We compared this proposed method with a benchmark method based on PDE techniques defined in \cite{slvKlebaner2015} with two different numerical examples. Under both cases, we have observed that the proposed method is more robust and has significantly smaller relative error when compared to the benchmark method.

Since our proposed method is aimed to improve the error created by using Equation (\ref{eq:update_L}) to updated the leverage function, we would have observed the same improvement documented in Section \ref{sec:conclusion} if we had used the adjoint method proposed in \cite{slv_numerical_pde_2017} to solve the related Fokker-Planck PDE.

Future development could consider the implementation of the online calibration procedure of \cite{AAZ2017}. This could not be achieved for the benchmark method. Another avenue would be to explore the fast mean reversion stochastic volatility setting conjoined with
the local volatility surface estimation as described in \cite{NP2006}.

\appendix
\section{Specification of the ADI method}\label{sec:app}

To solve Equation (\ref{eq:fokker_planck}) numerically, we apply the finite difference Douglas-Rachford (DR) method~\cite{douglas1956}.
For completeness, we shall now give the details of the implementation. 

We suppose $(t,S,V) \in [t_{\min},t_{\max}]\times [S_{\min},S_{\max}]\times [V_{\min},V_{\max}]$. The discretization contains $N_S+1$ nodes in $S$ direction, $N_V+1$ nodes in $v$ direction and $N_t+1$ nodes in $t$ direction. By using the central difference for the first-order differentiation, all partial differentiations could be approximated as follows:
\begin{align*}
\frac{\partial (Sp)}{\partial S} & \approx  \frac{S_{i+1}p_{n, i+1,j}-S_{i-1}p_{n, i-1,j}}{2\Delta S}=:\frac{\delta_S^{S} p_{n, i,j}}{2\Delta S}\nonumber\\
\frac{\partial \big((m-V)p\big)}{\partial V} & \approx  \frac{(m-V_{j+1})p_{n, i,j+1}-(m-V_{j-1})p_{n, i,j-1}}{2\Delta V}=:\frac{\delta_V^{m-V} p_{n, i,j}}{2\Delta V}\nonumber\\
\frac{\partial^2 \big(VL^2(t_n,S)S^2p\big)}{\partial S^2} & \approx  \frac{1}{(\Delta S)^2} (V_{j}L^2(t_n,S_{i+1})S_{i+1}^2p_{n, i+1,j}\\
&  -2V_{j}L^2(t_n,S_{i})S_{i}^2p_{n, i,j}+V_{j}L^2(t_n,S_{i-1})S_{i-1}^2p_{n, i-1,j})\nonumber\\
&=:  \frac{\delta_{SS}^{VL^2(t,S)S^2}p_{n, i,j}}{(\Delta S)^2}\nonumber\\
\frac{\partial^2 (Vp)}{\partial V^2} & \approx  \frac{V_{j+1}p_{n, i,j+1}-2V_{j}p_{n,i,j}+V_{j-1}p_{n,i,j-1}}{(\Delta V)^2}=:\frac{\delta_{VV}^{V}p_{n,i,j}}{(\Delta V)^2}\nonumber\\
\frac{\partial^2 \big(VL(t_n,S)Sp\big)}{\partial S \partial V} & \approx  \frac{1}{4\Delta S \Delta V}(V_{j+1}L(t_n,S_{i+1})S_{i+1}p_{n,i+1,j+1}+\nonumber\\
&  V_{j-1}L(t_n,S_{i-1})S_{i-1}p_{n,i-1,j-1}-V_{j+1}L(t_n,S_{i-1})S_{i-1}p_{n,i-1,j+1}-\nonumber\\
&  V_{j-1}L(t_n,S_{i+1})S_{i+1}p_{n,i+1,j-1})\nonumber\\
&=:   \frac{\delta_{SV}^{VL(t,S)S}p_{n,i,j}}{4\Delta S\Delta V}\nonumber
\end{align*} 

We replace the derivative in Equation (\ref{eq:fokker_planck}) by these finite difference quotients. We then define the discretized system for the approximation $p_{n, i,j}$ for $p(t_n, S_i, V_j)$ given by the $\theta$-scheme: 
\begin{equation}
(1-\theta A_1-\theta A_2) p^{(n+1)}=[1+A_0+(1-\theta)A_1+(1-\theta) A_2]p^{(n)} + O(\Delta t^3)\nonumber
\end{equation}
for $n=0,1,2,\ldots, N_t-1$, where $p^{(n)} = \{ p_{n,i,j}\}_{i,j=0}^{N_S, N_V}$, $\theta \in [0,1]$ and 
\begin{eqnarray}
A_0 & := & \frac{1}{4} \rho \xi R_{SV} \delta_{SV}^{VL(t,S)S}, \nonumber\\
A_1 & := & R_{S2}\delta_{SS}^{VL^2(t,S)S^2}+\frac{1}{2}(r-d)R_S \delta_S^S, \nonumber\\
A_2 & := & \xi^2 R_{V2}\delta_{VV}^V+\frac{1}{2}\kappa R_V \delta_V^{m-V}, \nonumber
\end{eqnarray}
$$R_S:=\frac{\Delta t}{\Delta S},\ R_V:=\frac{\Delta t}{\Delta V},\ R_{S2}:=\frac{\Delta t}{\Delta S^2},\ R_{V2}:=\frac{\Delta t}{\Delta V^2},\ R_{SV}:=\frac{\Delta t}{\Delta S\Delta V}.$$

The Douglas-Rachford method (DR method) is then defined as:
\begin{eqnarray}
(1-\theta A_1) W & = & [1+A_0+(1-\theta)A_1+A_2]p^{(n)}\nonumber\\
(1-\theta A_2) p^{(n+1)} & = & W-\theta A_2 p^{(n)}\nonumber
\end{eqnarray} 

Note that, here, for notational reason, we assume the rate $r-d$ is constant. In our experiment, with a slight modification of $A_1$ and $A_2$, we developed the method for the case of $r-d$ being time-dependent and also the zero flux condition\cite{boundary_flux} .

\bibliographystyle{alpha}
\bibliography{slv_calibration_saporito_yang_zubelli}


\end{document}